\documentclass[aps,english,superscriptaddress,preprintnumbers,reprint,footinbib,amsmath,amssymb,prb]{revtex4-2}
\usepackage[T1]{fontenc}
\usepackage[utf8]{inputenc}
\usepackage[version=3]{mhchem}
\usepackage{graphicx}
\usepackage{dcolumn}
\usepackage{bm}
\usepackage{hyperref}
\hypersetup{
    colorlinks=true,
    linkcolor=blue,
    filecolor=blue,
    urlcolor=blue,
   citecolor=blue,
}
\usepackage{siunitx}
\usepackage{enumitem}
\usepackage{dsfont}
\usepackage[dvipsnames]{xcolor}
\usepackage{times} 
\usepackage{newtxtext,newtxmath}
\usepackage{xcolor}
\usepackage{soul}

\begin{document}

\title{Qubit Gate Operations in Elliptically Trapped Polariton Condensates}

\author{Luciano S. Ricco}
\email{lsricco@hi.is}
\affiliation{Science Institute, University of Iceland, Dunhagi-3, IS-107 Reykjavik, Iceland}


\author{Ivan A. Shelykh}
\affiliation{Science Institute, University of Iceland, Dunhagi-3, IS-107 Reykjavik, Iceland}
\affiliation{Russian Quantum Center, Skolkovo IC, Bolshoy Bulvar 30 bld. 1, Moscow 121205, Russia}

\author{Alexey Kavokin}
\email{a.kavokin@westlake.edu.cn}
\affiliation{Key Laboratory for Quantum Materials of Zhejiang Province, School of Science, Westlake University, 310024 Hangzhou, China}
\affiliation{Institute of Natural Sciences, Westlake Institute for Advanced Study, 310024 Hangzhou, China}
\affiliation{Spin Optics Laboratory, St. Petersburg State University, 198504 St. Petersburg, Russia}

\date{\today}

\begin{abstract}
We consider bosonic condensates of exciton-polaritons optically confined in elliptical traps. A superposition of two non-degenerated \textit{p}-type states of the condensate oriented along the two main axes of the trap is represented by a point on a Bloch sphere, being considered as an optically tunable qubit. We describe a set of universal single-qubit gates resulting in a controllable shift of the Bloch vector by means of an auxiliary laser beam. Moreover, we consider interaction mechanisms between two neighboring traps that enable designing two-qubit operations such as CPHASE, \textit{i}SWAP, and CNOT gates. Both the single- and two-qubit gates are analyzed in the presence of error sources in the context of polariton traps, such as pure dephasing and spontaneous relaxation mechanisms, leading to a fidelity reduction of the final qubit states and quantum concurrence, as well as the increase of Von Neumann entropy. We also discuss the applicability of our qubit proposal in the context of DiVincenzo's criteria for the realization of local quantum computing processes. Altogether, the developed set of quantum operations would pave the way to the realization of a variety of quantum algorithms in a planar microcavity with a set of optically induced elliptical traps.
\end{abstract}

\maketitle

\section{Introduction}

The placement of solid-state exciton-polariton systems, here on just polaritons, in the quantum computing race remains questionable to date~\cite{KavokinNatRevPhys2022, Liew_OptMatExp2023} despite the growing quality of patterned optical cavities and available active materials therein~\cite{Sanvitto_NatMat2016, Keeling_AnnRev2020} and steadily advancing techniques in potential landscape engineering~\cite{Schneider2017RPP}, and optical control to minimize decoherence processes~\cite{Topfer_Optica2021}. In inorganic III-V semiconductor microcavities, significant cross-phase modulation~\cite{Kuriakose_NatPhot2022}, squeezing~\cite{Boulier_NatComm2014}, blockade effect~\cite{Delteil_NatMat2019, Munoz_NatMat2019} and interactions~\cite{Cuevas_SciAdv2018} at the single polariton level have already been demonstrated due to the large interaction strengths between polaritons, owing to the generous size of their underlying Wannier-Mott exciton component. Nowadays, quantum computing proposals using polaritons can be divided into two categories: single-particle~\cite{Kyriienko2016, Puri_PRB2017, Xu_PRB2021, Nigro_CommPhys2022} and macroscopic field~\cite{Demirchyan_PRL2014, Solnyshkov_SupLatt2015, Ghosh2020, Xue2021} strategies. Here, we are concerned with the latter based on nonequilibrium polariton Bose-Einstein condensates. 

Polaritons are hybrid particles arising in the strong coupling regime between matter (excitons) and light (confined photons)~\cite{Carusotto2013}. They possess extremely light effective mass, large inter-particle interaction strengths, and can be reversibly adjusted through all-optical techniques. Importantly, information about the polariton state is encoded in the emitted cavity light which can be measured through standard optical techniques.  Being bosonic quasiparticles, polaritons can be stimulated into a macroscopically coherent state which lies at the interface between nonequilibrium Bose-Einstein condensates and polariton lasers~\cite{Kasprzak2006}. In the mean-field picture, a condensate of polaritons can be conveniently described by a single macroscopic wavefunction $\Psi(\mathbf{r},t)$~\cite{Carusotto2013}. 

Because polariton condensates are driven-dissipative objects, with particles being generated from an external laser excitation and losses naturally occurring through the cavity mirrors, they can possess equilibrium points that do not coincide with the many-body system ground state in thermodynamic equilibrium. In particular, they can populate and stabilize into the excited state manifolds of their transverse trapping configuration which includes micropillars~\cite{Lukoshkin_PRB2018, Sedov_ACSPhot2020, Sedov2021, Real_PRR2021, Lukoshkin_SciRep2023}, patterned mesas~\cite{NardinPhysRevB82073303(2010), Gao_PRL2018}, cavities with metallic deposition~\cite{Kim_NatPhys2011}, and optically generated potentials~\cite{Askitopoulos2015, Dall_PRL2014, Sun_PRB2018, Askitopoulos_PRB2018, Topfer_PRB2020, Sitnik_PRL2022, Barrat2023superfluid}. In particular, these optically induced potentials can be flexibly designed with the use of spatial light modulators (SLMs). Their shape may be varied on demand from one experiment to another, permitting the realization of polariton \textit{XY}~\cite{Berloff2017, Harrison_PRAppl2022} or Ising simulators~\cite{Kalinin_Nanopho2020, alyatkin2022alloptical}.

The concept of using polaritons as macroscopic quantum states for continuous-variable quantum computation~\cite{Oliveira_PRA2000} was recently visited by Xue \textit{et al.}~\cite{Xue2021} using the superposition of co-localized and non-degenerate ring-shaped polariton condensates of opposite circulation. These polariton qubits are similar to their superconducting counterpart, the flux qubits. However, instead of circulating basis states, polariton condensates might be more conveniently described in terms of \textit{p}-states, corresponding to their spatial dipolar distribution along the two main axes of the transverse trap, i.e., $p_{x}$ and $p_{y}$. In the context of optical traps, the equal linear combination of $p_{x}$ and $p_{y}$ states corresponds to polariton condensates with an integer orbital angular momenta (OAM), whose the \textit{p}-states wavefunctions are real and characterized by odd parity. In an ideal round trap potential, these $p_x$ and $p_y$ modes are degenerate, however, any small geometric ellipticity induces an energy splitting between them~\cite{Askitopoulos2015,  GaelNardinJournalofNanophot2011, Bennenhei_OptMaterExpress2023}.
 
 For the purpose of the realization of digital quantum computers based on polariton qubits~\cite{KavokinNatRevPhys2022}, one should aim at maximizing the coherence times of each individual qubit and coupled arrays of qubits. Interestingly, the dynamics of polariton condensates are characterized by a range of time scales, and it is a non-trivial question of which of them would be responsible for the decoherence of polariton qubits. The shortest timescale is given by the single polariton lifetime that is dependent on the quality factor of the cavity and can reach hundreds of ps in planar cavities~\cite{Nelsen_PRX2013} and grated waveguides utilizing photonic states protected from the continuum~\cite{Ardizzone_Nature2022}. The coherence time of a polariton laser condensate, measured by the time-resolved interferometry measurements, can be two orders of magnitude longer, (\textit{i.e.}, several ns), because of the stimulated scattering of polaritons that stabilizes their final state~\cite{Caputo_NatMat2018, Askitopoulos_Arxiv2019, Sigurdsson_PRL2022}. Moreover, the spatial coherence of an optically trapped continuous wave (CW)-driven polariton condensate has been measured to be practically uniform within the trap~\cite{Topfer_PRB2020, Sigurdsson_PRL2022} underlining the condensate's quality as a macroscopic spatially coherent object. The ultimate upper bound is the duration of the laser excitation which sustains the condensates which can be in the range of milliseconds before sample heating becomes a problem.

  \begin{figure}[t]
\centerline{\includegraphics[width=0.49\textwidth,keepaspectratio]{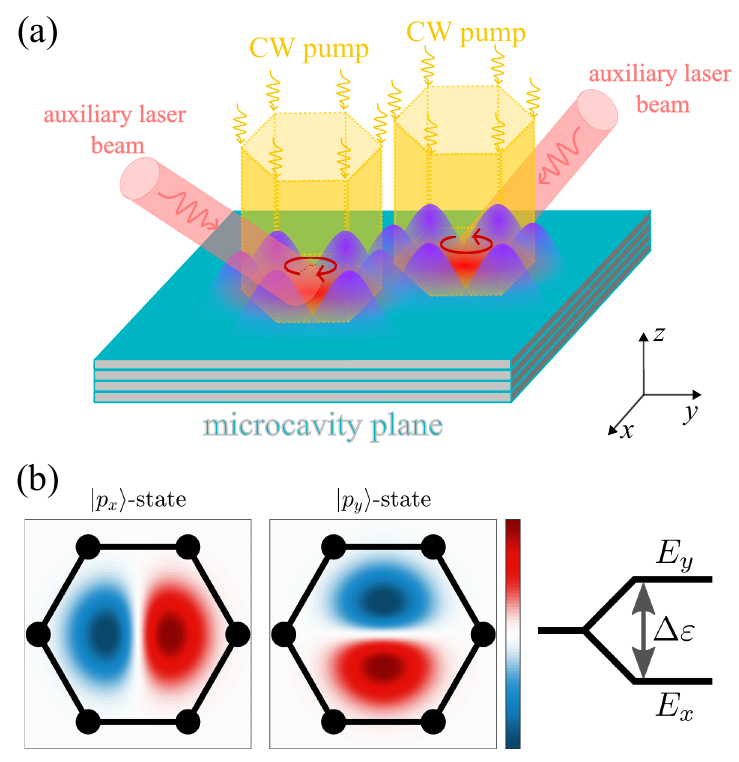}}
\caption{\label{Fig:Device}(a) Sketch of the proposed polariton condensate qubit system. Two polariton condensates are formed at distinct regions of a microcavity plane, within an elliptical trap in the presence of a nonresonant continuous wave (CW) laser pump with a hexagonal shape. These trapped condensates host superfluid circularly current states whose direction is defined by opposite OAM, as represented by the red circled arrows in each trap. The linear combination of clockwise and anticlockwise OAM states defines the orthogonal $|p_x\rangle$- and $|p_y\rangle$-modes in the optically generated trap and are the basis states for a two-level qubit in each polariton condensate. Both the initial qubit states and qubit gates for an individual trap can be tuned by the corresponding auxiliary laser beams on resonance with the microcavity photonic modes. Moreover, the coupling between two qubits can be switched on and off by a control laser pulse that creates a potential barrier between the two traps (not shown). 
(b) Spatial representation of the \textit{p}-modes, corresponding to the qubit basis states. The right panel illustrates the energy splitting $\Delta \varepsilon$ between the modes, which can be tuned by stretching out the example hexagonal trap.
}
\end{figure}

Here, we consider a macroscopic polariton-based qubit composed of non-degenerate $|p_{x,y}\rangle$-states confined within an optically induced two-dimensional trap, as illustrated in Fig.~\ref{Fig:Device}. Specifically, these basis states comprise the $|p_x\rangle$- and $|p_y\rangle$-modes, corresponding to the spatial dipolar distribution along the two principal axes of the trap. The energy difference between these states is determined by the ellipticity of the trap, which can be adjusted using an SLM to shape the laser responsible for the generation of the trap potential~\cite{Topfer_PRB2020, Gnusov_PRAppl2021}. By manipulating the trapped polariton condensate with an auxiliary resonant laser beam, we establish a theoretical framework for a universal set of single-qubit operations. These operations allow for precise control over the qubit state vector within the Bloch sphere using purely optical means. Additionally, we explore a set of two-qubit gates, enabled by distinct coupling mechanisms between the $|p_x\rangle$- and $|p_y\rangle$-states of neighboring traps. These interactions can be selectively blocked through nonresonant laser control pulses that create an effective potential barrier between the traps~\cite{Alyatkin_PhysRevLett124207402(2020)}. In particular, by tuning the parameters of the auxiliary laser beams, their operational duration, and the interaction between the traps, we define essential two-qubit operations for quantum computing processes. These include fundamental gates such as the $\imath$SWAP, CPHASE, and CNOT operations, with the latter two falling into the category of \textit{entangling gates}~\cite{KrantzetalQuantumEnginnersGuideAppPhysRev2019}. Furthermore, we assess the fidelity of the output state for representative quantum gates, e.g., Hadamard and CPHASE, along with key metrics such as Von Neumann entropy and quantum concurrence. These analyses take into account the presence of intrinsic error sources, namely pure dephasing and spontaneous relaxation, which can affect the performance of polariton-based qubits. Lastly, we briefly examine how our proposal of qubits based on optically trapped polariton condensates aligns with each of the five DiVincenzo's criteria~\cite{DiVincenzo2000,Book_nielsen_chuang_2010} for the practical implementation of local quantum computing processes. 

 
The polariton condensate qubit system hereby explored presents some notable advantages when compared to qubits based on split-ring polariton condensates proposed in Ref.~\cite{Xue2021}. One key advantage is the ability to achieve direct optical control over the energy splitting between the two polariton states in the system by reprogramming the SLM to adjust the eccentricity of the optical trap~\cite{Gnusov_PRAppl2021}. This control directly impacts the period of Rabi oscillations and, consequently, the characteristic time required to execute a single-qubit quantum operation~\cite{Barrat2023superfluid}. We also highlight that the present scheme differs significantly from the proposal by Ref.~\cite{Ghosh2020}, which relies on a single-polariton nonlinearity and the polariton blockade effect giving rise to a qubit basis built upon quantum fluctuations on top of polariton condensates within semiconductor micropillars. Moreover, the macroscopic polariton qubit explored here offers the unique capability to optically switch on/off the interaction between neighboring traps, or equivalently, between the two qubits. This feature provides a natural advantage compared to superconducting-based qubits, where interaction cannot be completely excluded. 
In contrast, one main concern present is error correction for polaritons which would need at least some form of active feedback mechanism that fixes errors on GHz polariton rates whereas most SLM technologies are operating at KHz frequencies with few exotic solutions pushing the GHz limit~\cite{Peng_OptExpr2019}. Still, the unprecedented rapid development of polariton computing might enable one to circumvent this shortcoming by repeating the computation for a sufficient amount of time in order to accumulate the statistics of results, allowing one to filter out the correct solution.

\section{Model Hamiltonian and Quantum Gate Operations}

\subsection{Single-qubit gate operations}\label{SingleQubitGates}

We start by defining canonical single qubit operations~\cite{KrantzetalQuantumEnginnersGuideAppPhysRev2019} on a single trapped condensate and later move onto defining quantum gate operations between two trapped polariton condensates (i.e., a two-qubit system). Single qubit gates are mathematically defined as unitary rotations of the qubit state around a given axis of the Bloch sphere~\cite{Book_nielsen_chuang_2010} which necessitates a suitable basis of quantum states to work with. A good basis might rely on the symmetric and antisymmetric polariton levels~\cite{Demirchyan_PRL2014}, the natural two-component spin structure of polaritons~\cite{Solnyshkov_SupLatt2015}, or counterrotating superfluid currents~\cite{Sedov_arxiv2019, Xue2021, Barrat2023superfluid}. Instead, we define our basis using the orthogonal $|p_x\rangle$ and $|p_y\rangle$ states of an optically generated trap~\cite{Askitopoulos_PRB2018, Cherotchenko_PRB2021} whose energy levels can be continuously split, $\Delta \varepsilon = E_x - E_y$, by simply adjusting the nonresonant excitation beam profile into an elliptical annular shape. Our choice of basis is motivated by the recent demonstrations of spatially coupled polariton condensate vortices~\cite{Ma_NatComm2020, alyatkin2022alloptical} whose coherent superposition of clockwise $|\circlearrowright\rangle$ and anticlockwise $|\circlearrowleft\rangle$ OAM defines the trap's dipolar states, 
\begin{equation}
\begin{split}
|p_{x}\rangle&=\frac{1}{\sqrt{2}}\left(|\circlearrowright\rangle+|\circlearrowleft\rangle\right), \\
|p_{y}\rangle&=\frac{1}{\sqrt{2}}\left(|\circlearrowright\rangle-|\circlearrowleft\rangle\right), 
\end{split} \label{eq: px,py}
\end{equation}
 which are used as building blocks to define a qubit basis, i.e., $|p_{x}\rangle \equiv |0\rangle$ and $|p_{y}\rangle \equiv |1\rangle$. We assume that the system is pumped only slightly above the condensation threshold in order to avoid fragmentation of the condensate across multiple trap modes and power-induced collapse into the $|s\rangle$ ground state~\cite{Topfer_PRB2020}. The convenient feature of this basis is the spatial structure of the basis states which permits resonant excitation of different kinds of superpositions of $|p_{x}\rangle$ and $|p_{y}\rangle$~\cite{GaelNardinJournalofNanophot2011,NardinPhysRevB82073303(2010),CernaPhysRevB80121309(2009),Ma_NatComm2020}. Within the two-level qubit subspace $\{|0\rangle,|1\rangle\}$, setting $\hbar = 1$, the effective single-qubit Hamiltonian can be written: 
\begin{equation}
\hat{\mathcal{H}} =  \mathcal{P}_x\hat{\sigma}_{x} + \mathcal{P}_y\hat{\sigma}_{y} + \frac{\Delta \varepsilon}{2}\hat{\sigma}_{z}, \label{eq:single_qubit}
\end{equation}
The operators $\hat{\sigma}_{x,y,z}$ are the standard Pauli matrices, and can be written in terms of the qubit basis as $\hat{\sigma}_{z} = |0\rangle\langle0|-|1\rangle\langle1|$, $\hat{\sigma}_{x} = |0\rangle\langle1|+|1\rangle\langle0|$ and $\hat{\sigma}_{y} = -\imath(|0\rangle\langle1|-|1\rangle\langle0|)$, with $|0\rangle = \begin{pmatrix}1 & 0\end{pmatrix}^{T}$ and $|1\rangle = \begin{pmatrix}0 & 1\end{pmatrix}^{T}$.  Thus, the $\mathcal{P}_x,\mathcal{P}_y$ parameters are responsible for changing the coupling between $|0\rangle$ ($|p_x\rangle$) and $|1\rangle$ ($|p_y\rangle$) states, while $\Delta \varepsilon$ is the energy splitting. 

Control over the parameters $(\mathcal{P}_x, \mathcal{P}_y, \Delta \varepsilon)$ comes from a suite of SLM techniques that change the shape and form of the nonresonant pump inducing the optical trap, and subsequently the coupling between the $|p_x\rangle$ and $|p_y\rangle$ states as demonstrated in Ref.~\cite{Barrat2023superfluid}. The last term is proportional to the eccentricity of the trap which affects the splitting along the minor and major axis~\cite{Askitopoulos2015, Topfer_PRB2020}. The second term comes from adjusting the in-plane orientation of the trap~\cite{Gnusov_PRAppl2021}; e.g. a $45^\circ$ degree rotation couples $|p_x\rangle$ and $|p_y\rangle$ to form their diagonal and antidiagonal counterparts. The first term couples the $|p_x\rangle$ and $|p_y\rangle$ states to form circulating currents which can be achieved by setting the trap into rotational motion~\cite{Gnusov_SciAdv2023, delValle_NanoLetters_2023}. The condensate population can also be tuned gradually from the excited state to the ground state, i.e., sweep the polar angle of the Bloch sphere, with pump power~\cite{Topfer_PRB2020}. From here on, we will refer to these control inputs as nonresonant {\it auxiliary laser} inputs as indicated in Fig.~\ref{Fig:Device}.

Following the definition expressed in Eq.~(\ref{eq: px,py}), the single qubit Hamiltonian of Eq.~(\ref{eq:single_qubit}) can be written in its corresponding OAM basis $\{|\circlearrowright\rangle,|\circlearrowleft\rangle\}$ as follows:
\begin{eqnarray}
    \hat{\mathcal{H}} & = & \mathcal{P}_x(|\circlearrowright\rangle\langle\circlearrowright|-|\circlearrowleft\rangle\langle\circlearrowleft|)\nonumber\\
    & + &\left[\left(\frac{\Delta\varepsilon}{2} + \imath \mathcal{P}_y\right)|\circlearrowright\rangle\langle\circlearrowleft| + \text{h.c.}\right]. \label{eq:H_Qubit_VortexBasis1}
\end{eqnarray}
In the OAM subspace, we can notice that the \textit{x}-component of the auxiliary laser shifts the energy splitting of the clockwise and anticlockwise OAM states. Simultaneously, both the \textit{z}-component and \textit{y}-component modulate the coupling between OAM states with opposing directions when confined within the same trap. In particular, Barrat {\it et al.}~\cite{Barrat2023superfluid} achieve this manipulation of the interactions between counter-circulating polariton trap states by using an auxiliary laser in the form of an off-centered Gaussian "bump".
     

For an operational time $\tau$ of the auxiliary laser beam, the temporal evolution of the single qubit Hamiltonian expressed in Eq.~(\ref{eq:single_qubit}) is given by the unitary operator $\hat{\mathcal{U}}(\tau)=e^{-\imath \hat{\mathcal{H}}\tau}$~\cite{Book_sakurai_napolitano_2017} so that the final qubit state becomes $|\psi\rangle = \hat{\mathcal{U}}(\tau) |\psi_{0}\rangle$, for a given initial qubit state $|\psi_{0}\rangle$. To explore the effect of such unitary time evolution, the single qubit Hamiltonian is expressed in a Bloch sphere representation~\cite{KrantzetalQuantumEnginnersGuideAppPhysRev2019,Book_nielsen_chuang_2010}, by considering a parameterized unit vector $\hat{\boldsymbol{n}}$ in spherical coordinates, in which $\hat{\boldsymbol{n}}\in \mathbb{R}^{3}$ is called Bloch vector, where the phase $\theta$ of the laser beam is the polar angle of the Bloch sphere ($0\leq\theta\leq\pi$), with a parameterized azimuthal angle $\phi$ ($0\leq\phi<2\pi$). In this way, $\hat{\boldsymbol{n}}=(\cos{\theta}\sin{\phi},\sin{\theta}\sin{\phi},\cos{\phi})$, so that 
\begin{equation}
 \hat{\boldsymbol{n}} = \frac{\boldsymbol{\mathcal{P}}}{|\boldsymbol{\mathcal{P}}|},\label{eq:UnitVector} 
\end{equation}
where $\boldsymbol{\mathcal{P}}=\left(\mathcal{P}_{0}\cos{\theta},\mathcal{P}_{0}\sin{\theta}, \frac{\Delta\varepsilon}{2} \right)$, with the norm $|\boldsymbol{\mathcal{P}}|\equiv\mathcal{P}= \sqrt{\mathcal{P}_{0}^{2} + \frac{\Delta\varepsilon^{2}}{4}}$. From this parameterization, $\phi = \arccos{(\frac{\Delta\varepsilon}{2\mathcal{P}})}$. 
The single-qubit Hamiltonian can then be rewritten as,
\begin{equation}
\hat{\mathcal{H}} = \mathcal{P}\boldsymbol{\sigma}\cdot\hat{\boldsymbol{n}},\label{eq:H_Sigle_Parametrized}   
\end{equation}
with the corresponding unitary operator given by~\cite{Ghosh2020}:
\begin{align} \notag
&\hat{\mathcal{U}}(\mathcal{P}\tau)  = \\
& \begin{bmatrix}\cos(\mathcal{P}\tau)-\imath\cos\phi\sin(|\boldsymbol{\mathcal{P}}|\tau) & -\imath e^{-\imath\theta}\sin\phi\sin(\mathcal{P}\tau)\\
-\imath e^{\imath\theta}\sin\phi\sin(\mathcal{P}\tau) & \cos(\mathcal{P}\tau)+\imath\cos\phi\sin(\mathcal{P}\tau)
\end{bmatrix}.\label{eq:UnitaryOp_SingleQubit}
\end{align}
%
The unitary operator in Eq.~(\ref{eq:UnitaryOp_SingleQubit}) represents distinct single-qubit gates by tuning the auxiliary laser beam parameters during a time interval $\tau$. Let us consider the following cases, for instance:
\begin{equation}
\hat{\mathcal{U}}\left(\tau=\frac{\pi}{2\mathcal{P}},\theta=0,\phi=\frac{\pi}{2}\right)=\begin{bmatrix}0 & -\imath\\
-\imath & 0
\end{bmatrix}\equiv e^{-\imath\frac{\pi}{2}}\hat{X}_{\pi}, \label{eq:X_pi}   
\end{equation}
\begin{equation}
\hat{\mathcal{U}}\left(\tau=\frac{\pi}{2\mathcal{P}},\theta=\phi=\frac{\pi}{2}\right)=e^{-\imath\frac{\pi}{2}}\begin{bmatrix}0 & -\imath\\
\imath & 0
\end{bmatrix}\equiv e^{-\imath\frac{\pi}{2}}\hat{Y}_{\pi}, \label{eq:Y_pi}   
\end{equation}
and
\begin{equation}
\hat{\mathcal{U}}\left(\tau=\frac{\pi}{2\mathcal{P}},\theta=\forall,\phi=\pi\right)=\begin{bmatrix}\imath & 0\\
0 & -\imath
\end{bmatrix}\equiv e^{\imath\frac{\pi}{2}}\hat{Z}_{\pi}, \label{eq:Z_pi}   
\end{equation}
where $\hat{X}_{\pi}$, $\hat{Y}_{\pi}$ and $\hat{Z}_{\pi}$ are known as Pauli gates~\cite{Book_nielsen_chuang_2010,KrantzetalQuantumEnginnersGuideAppPhysRev2019}, which rotates the qubit state by $\pi$ radians around \textit{x-}, \textit{y-} and \textit{z-}axis, respectively. Thus, by manipulating the auxiliary laser beam parameters, within a specified operational time $\tau$, we can effectively implement all the Pauli gates. These single-qubit operations belong to a universal quantum gate set denoted as $\mathcal{G}_{0}=\{\hat{X}_{\varphi},\hat{Y}_{\varphi},\hat{Z}_{\varphi},\text{Ph}_{\varphi}, \text{CNOT} \}$~\cite{KrantzetalQuantumEnginnersGuideAppPhysRev2019}. The CNOT operation, a two-qubit gate, will be defined in the subsequent section. 

In the context of quantum computing operations, another crucial single-qubit gate is the so-called Hadamard gate $\hat{H}$~\cite{Book_nielsen_chuang_2010,KrantzetalQuantumEnginnersGuideAppPhysRev2019}. The Hadamard gate is responsible for generating an equal superposition of states, forming the foundation of qubit basis manipulation. In the framework of our proposal, we can derive this gate from the unitary operator defined in Eq.~\eqref{eq:UnitaryOp_SingleQubit}, as follows:
\begin{equation}
\hat{\mathcal{U}}\left(\tau=\frac{\pi}{2\mathcal{P}},\theta=0,\phi=\frac{\pi}{4}\right)=\frac{-\imath}{\sqrt{2}}\begin{bmatrix}1 & 1\\
1 & -1
\end{bmatrix}\equiv e^{-\imath\frac{\pi}{2}}\hat{H},\label{eq:Hadamard}  
\end{equation}
which represents the standard Hadamard gate, accompanied by an overall phase gate $\text{Ph}_{-\frac{\pi}{2}}=e^{-\imath\frac{\pi}{2}}\mathds{1}$ acting on the qubit state.

To illustrate the practical effect of the Hadamard gate as defined in Eq.~\eqref{eq:Hadamard}, let us assume that the elliptical trap is initially set to favor the state $|p_{x}\rangle \equiv |0\rangle$. Consequently, we have:
\begin{equation}
e^{-\imath\frac{\pi}{2}}\hat{H}|0\rangle = \frac{e^{-\imath\frac{\pi}{2}}}{\sqrt{2}}\left(|0\rangle+|1\rangle\right)\equiv \text{Ph}_{-\frac{\pi}{2}}|\circlearrowright\rangle. \label{eq:Hadamard_to_|zero>}  
\end{equation}
It is evident that the Hadamard gate operation on the initial state $|0\rangle$ results in a clockwise OAM state [see Eq.~(\ref{eq: px,py})]. Similarly, if the trap's initial state is set to $|p_{y}\rangle \equiv |1\rangle$, after applying Eq.~(\ref{eq:Hadamard}), the final single-qubit state will be the anticlockwise OAM state $|\circlearrowleft\rangle$. This demonstrates that the Hadamard gate operation manipulates the direction of OAM states within the trap, depending on the qubit initial state.

\subsection{Two-qubit gate operations}\label{TwoQubitGates}
%
We next introduce a two-qubit Hamiltonian, given by:
\begin{equation}
 \hat{\mathcal{H}}_{12}=\sum_{j=1,2}\left(\mathcal{P}_{x}^{(j)}\hat{\sigma}_{x}^{(j)}+\mathcal{P}_{y}^{(j)}\hat{\sigma}_{y}^{(j)} + \frac{\Delta\varepsilon_{j}}{2}\hat{\sigma}_{z}^{(j)} \right) + \hat{\mathcal{H}}_{\text{int}}, \label{eq:H_twoqubit}   
\end{equation}
where $\hat{\mathcal{H}}_{\text{int}}$ accounts for the interaction between the two qubits, or equivalently, between the macroscopic dipolar modes of distinct traps. 
The operators for each qubit are expanded within their respective subspaces, denoted as $\hat{\sigma}_{x,y,z}^{(1)}=\hat{\sigma}_{x,y,z}\otimes\mathds{1}$ and $\hat{\sigma}_{x,y,z}^{(2)}=\mathds{1}\otimes\hat{\sigma}_{x,y,z}$, where $\mathds{1}$ represents a $2\times2$ identity matrix.

The Hamiltonian~\eqref{eq:H_twoqubit} indicates that, in addition to the parameters of the auxiliary laser beam, two-qubit gate operations are influenced by the interaction between the individual qubits, specifically, the interaction between neighboring traps. Typically, this interaction is represented as $\hat{\mathcal{H}}_{\text{int}}=J_{k}(\hat{\sigma}_{k}\otimes\hat{\sigma}_{k})$, where $k=x,y,z$, and $J_{k}$ represents the strength of the qubit-qubit interaction. Within our proposal, the coupling between adjacent qubits can be adjusted using a control laser pulse, which creates an all-optical potential barrier between the elliptical traps~\cite{alyatkin2022alloptical}, or through the use of an acousto-optic modulator~\cite{Barrat2023superfluid}.

For instance, for an Ising-type interaction in the qubit basis, $J_{x}=J_{y}=0$ and $J_{z}=J_{12}$, which leads to the following interacting Hamiltonian:
\begin{equation}
 \hat{\mathcal{H}}_{\text{int}} = J_{12}(\hat{\sigma}_{z}^{(1)}\cdot \hat{\sigma}_{z}^{(2)})=J_{12}(\hat{\sigma}_{z}\otimes \hat{\sigma}_{z}). \label{eq:H_int_Ising_type} 
\end{equation}
By considering this interaction into Eq.~(\ref{eq:H_twoqubit}), and tuning the system parameters so that $\Delta\varepsilon_{1}=\Delta\varepsilon_{2}=-2J_{12}$ and $\mathcal{P}_x^{(j)},\mathcal{P}_{y}^{(j)}\ll J_{12}$ (weak auxiliary laser beam), the two-qubit Hamiltonian is reduced to:
\begin{equation}
\hat{\mathcal{H}}_{12}=J_{12}\left(\hat{\sigma}_{z}\otimes\hat{\sigma}_{z}-\hat{\sigma}_{z}\otimes\mathds{1}-\mathds{1}\otimes\hat{\sigma}_{z}\right).\label{eq:H_two_qubit_CPHASE}   
\end{equation}
The unitary operator that describes the time-evolution of Eq.~(\ref{eq:H_two_qubit_CPHASE}) for a given time duration $\tau$ of the auxiliary laser beams reads:
\begin{equation}
\hat{\mathcal{U}}_{12}(J_{1,2},\tau)=e^{\imath J_{12}\tau}\begin{bmatrix}1 & 0 & 0 & 0\\
0 & 1 & 0 & 0\\
0 & 0 & 1 & 0\\
0 & 0 & 0 & e^{-\imath4J_{12}\tau}
\end{bmatrix}. \label{eq:U_12_CPHASE}
\end{equation}

By setting $J_{1,2}\tau=\frac{\pi}{4}$ in the neighboring traps, such a unitary operator simplifies to:
\begin{equation}
\hat{\mathcal{U}}_{12}\left(\frac{\pi}{4} \right) = e^{\imath \frac{\pi}{4}} \text{CPHASE},\label{eq:CPHASE}
\end{equation}
where the term CPHASE represents a two-qubit gate operation. This operation applies a $\hat{\sigma}_z$ operation to the target qubit exclusively when the control qubit is in the state $|1\rangle$~\cite{Book_nielsen_chuang_2010}. The CPHASE gate falls under the category of \textit{entangling gates}, as well as the CNOT gate. This categorization arises from their capability to transform separate input states into entangled output states. Moreover, the application of a CPHASE gate in conjunction with two Hadamard gates [Eq.~(\ref{eq:Hadamard})] generates a CNOT gate, i.e, $\text{CNOT} =(\mathbb{I}\otimes\text{H})\text{CPHASE}(\mathbb{I}\otimes\text{H})$~\cite{KrantzetalQuantumEnginnersGuideAppPhysRev2019}.

As an illustrative example of applying a CPHASE gate to generate an entangled two-qubit state, let us begin with the initial state $|\psi_{0}\rangle$, where the optical traps are prepared with opposite OAM states, i.e., $|\psi_{0}\rangle =|\circlearrowright\rangle_{\text{C}}\otimes|\circlearrowleft\rangle_{\text{T}}=|\circlearrowright\rangle_{\text{C}}|\circlearrowleft\rangle_{\text{T}}$, with the subscripts C and T denoting the control and target qubits, respectively. Upon applying the CPHASE operation defined in Eq.~(\ref{eq:CPHASE}) to $|\psi_{0}\rangle$, we obtain the following result:
\begin{eqnarray}
 \hat{\mathcal{U}}_{12}\left(\frac{\pi}{4}\right)|\psi_{0}\rangle & = & \frac{e^{\imath\frac{\pi}{4}}}{2}\Big(|0\rangle_{\text{C}}|0\rangle_{\text{T}}-|0\rangle_{\text{C}}|1\rangle_{\text{T}}+|1\rangle_{\text{C}}|0\rangle_{\text{T}}\nonumber \\
 & + &|1\rangle_{\text{C}}|1\rangle_{\text{T}}\Big) \neq(\ldots)_{\text{C}}(\ldots)_{\text{T}},  \label{eq:CPHASE_final} 
\end{eqnarray}
in which the inequality indicates that the final two-qubit state cannot be factored into individual qubit subspaces, indicating the presence of entanglement between the qubits~\cite{Book_nielsen_chuang_2010,Book_sakurai_napolitano_2017}. We will see later that this entangled state leads to a maximal value of quantum concurrence. 

Another kind of two-qubit operation that can be implemented in our elliptically trapped polariton condensate by means of tuning the parameters of the auxiliary laser beans and interaction between the traps is the so-called \textit{i}SWAP gate~\cite{KrantzetalQuantumEnginnersGuideAppPhysRev2019}. This gate specifically requires an \textit{XY}-type interaction between the qubits, i.e, $J_{x}=J_{y}=J_{12}$ and $J_{z}=0$, which corresponds to,
\begin{equation}
\hat{\mathcal{H}}_{\text{int}} = J_{12}(\hat{\sigma}_{x}\otimes\hat{\sigma}_{x} + \hat{\sigma}_{y}\otimes\hat{\sigma}_{y}).\label{eq:H_int_XY}   
\end{equation}
By considering a trap of zero eccentricity so that $\Delta \varepsilon_{j}=0$, and also weak auxiliary laser beam $\mathcal{P}_x^{(j)},\mathcal{P}_y^{(j)}\ll J_{12}$, the total two-qubit Hamiltonian of Eq.~(\ref{eq:H_twoqubit}) is reduced only to the Hamiltonian describing the \textit{XY}-interaction in the qubit basis, cf. Eq.~(\ref{eq:H_int_XY}), with the corresponding unitary time-evolution operator:    
\begin{equation}
\hat{\mathcal{U}}_{12}(J_{1,2},\tau)= \left[
\begin{array}{cccc}
 1 & 0 & 0 & 0 \\
 0 & \cos (2 J_{12} \tau) & -\imath \sin (2 J_{12} \tau) & 0 \\
 0 & -\imath \sin (2 J_{12} \tau) & \cos (2 J_{12} \tau) & 0 \\
 0 & 0 & 0 & 1 \\
\end{array}
\right]. \label{eq:U_iSWAP}
\end{equation}
For $J_{12}\tau = \frac{\pi}{4}$, the unitary operator above corresponds exactly to the \textit{i}SWAP gate as follows:
\begin{equation}
\hat{\mathcal{U}}_{12}\left(\frac{\pi}{4}\right)= \left[
\begin{array}{cccc}
 1 & 0 & 0 & 0 \\
 0 & 0 & -\imath  & 0 \\
 0 & -\imath  & 0 & 0 \\
 0 & 0 & 0 & 1 \\
\end{array}
\right]\equiv i\text{SWAP}. \label{eq:iSWAP}
\end{equation}
The \textit{i}SWAP gate operation performs a state swap on the two-qubit system while introducing a phase difference of $\pi/2$. In practical terms, this means that in an illustrative scenario where the initial state is represented as $|\psi_{0}\rangle = |p_{x}\rangle_{\text{C}}|p_{y}\rangle_{\text{T}} \equiv |0\rangle_{\text{C}}|1\rangle_{\text{T}}$, the application of the \textit{i}SWAP gate, as defined in Eq.~(\ref{eq:iSWAP}), transforms it into $e^{-\imath\frac{\pi}{2}}|p_{y}\rangle_{\text{C}}|p_{x}\rangle_{\text{T}} \equiv e^{-\imath\frac{\pi}{2}}|1\rangle_{\text{C}}|0\rangle_{\text{T}}$ as the final state of the two-qubit system. 
Notice that both the CPHASE and \textit{i}SWAP gates, as previously defined, need the condition $J_{12}\tau=\frac{\pi}{4}$, along with the presence of weak auxiliary laser beams. However, the distinction between performing the CPHASE and the \textit{i}SWAP operation hinges on the adjustability of the ellipticity parameter $\Delta\varepsilon_j$.

The CNOT gate, which plays a role in entangling distinct qubit states~\cite{Book_nielsen_chuang_2010,KrantzetalQuantumEnginnersGuideAppPhysRev2019}, also can be implemented within our proposal. Unlike the previously defined CPHASE and \textit{i}SWAP gates, the CNOT operation requires individual control over the laser beam parameters for each trap. 
To demonstrate the CNOT gate implementation in the proposed device, we first initialize the two-qubit system with a CPHASE gate, allowing us to establish a two-qubit state basis comprising $\{|0\rangle_{\text{C}}|0\rangle_{\text{T}}, |0\rangle_{\text{C}}|1\rangle_{\text{T}}, |1\rangle_{\text{C}}|0\rangle_{\text{T}}, |1\rangle_{\text{C}}|1\rangle_{\text{T}}\}$. Subsequently, considering this basis, the unitary operator responsible for performing individual unitary operations $\hat{\mathcal{U}}{j}(\mathcal{P}{j},\tau_{j})$ on the \textit{jth}-qubit, as defined by Eq.~\eqref{eq:UnitaryOp_SingleQubit}, is expressed as follows:
\begin{equation}
\hat{\mathcal{U}}_{1|2}(\mathcal{P}_{1},\tau_{1};\mathcal{P}_{2},\tau_{2})=\begin{bmatrix}\hat{\mathcal{U}}_{1}(\mathcal{P}_{1},\tau_{1}) & 0\\
0 & \hat{\mathcal{U}}_{2}(\mathcal{P}_{2},\tau_{2})
\end{bmatrix},\label{eq:U_12_CNOT}
\end{equation}
 where $\mathcal{P}_{1,2}$ and $\tau_{1,2}$ are the parameterized norm [Eq.~(\ref{eq:UnitVector})] and time duration of the auxiliary laser beam in trap 1 and 2, respectively. Setting $\theta=0$ and using distinct laser operational times $\mathcal{P}_{1}\tau_{1}=\pi$ and $\mathcal{P}_{2}\tau_{2}=\frac{\pi}{2}$, the operator of Eq.~(\ref{eq:U_12_CNOT}) is reduced to:
 \begin{equation}
 \hat{\mathcal{U}}_{1|2}\left(\pi,\frac{\pi}{2}\right)=\begin{bmatrix}-1 & 0 & 0 & 0\\
0 & -1 & 0 & 0\\
0 & 0 & -\imath\cos\phi_{2} & -\imath\sin\phi_{2}\\
0 & 0 & -\imath\sin\phi_{2} & \imath\cos\phi_{2}
\end{bmatrix},     
\end{equation}
which is equivalent to a $-\text{CNOT}$ gate for $\phi_{2}=\frac{\pi}{2}$, with a phase of $\frac{\pi}{2}$ in the second qubit, i.e:
\begin{equation}
 \hat{\mathcal{U}}_{1|2}\left(\pi,\frac{\pi}{2}\right)\equiv - \begin{bmatrix}1 & 0 & 0 & 0\\
0 & 1 & 0 & 0\\
0 & 0 & 0 & \imath\\
0 & 0 & \imath & 0
\end{bmatrix}. \label{eq:-CNOT + phase}
 \end{equation}

A standard CNOT operation flips the target qubit if and only if the control bit is in the $|1\rangle$ state, as explained in~\cite{Book_nielsen_chuang_2010}. To illustrate this in our system, consider that the first trap (C-qubit) is configured to be in the $|\circlearrowright\rangle$ state, while the second trap (T-qubit) is appropriately excited to be in the $|p_{y}\rangle \equiv |1\rangle$ state. This sets up an initial state of $|\psi_{0}\rangle = |\circlearrowright\rangle_{\text{C}}|1\rangle_{\text{T}}$. When we apply the CPHASE operation, as defined in Eq.~(\ref{eq:CPHASE}), to this two-qubit state, it transforms into the following state:
\begin{equation}
|\psi\rangle = \frac{e^{\imath\frac{\pi}{4}}}{\sqrt{2}}\left(|0\rangle_{\text{C}}|1\rangle_{\text{T}}-|1\rangle_{\text{C}}|1\rangle_{\text{T}}\right) = e^{\imath\frac{\pi}{4}}|\circlearrowleft\rangle_{\text{C}}|1\rangle_{\text{T}}.\label{eq:Psi_before_CNOT}
\end{equation}
Notice that this final composite state of two qubits comprises separable states, thereby indicating an absence of entanglement. Now, by applying a CNOT gate [Eq.~(\ref{eq:-CNOT + phase})] on $|\psi\rangle$, one gets:
\begin{eqnarray}
 \hat{\mathcal{U}}_{1|2}|\psi\rangle & = & -\frac{e^{\imath\frac{\pi}{4}}}{\sqrt{2}}\left(|0\rangle_{\text{C}}|1\rangle_{\text{T}}-\imath|1\rangle_{\text{C}}|0\rangle_{\text{T}}\right)\nonumber\\
 &\neq &-e^{\imath\frac{\pi}{4}}\left(\ldots\right)_{\text{C}}\left(\ldots\right)_{\text{T}},\label{eq:Psi_after_CNOT}
 \end{eqnarray}
 which therefore leads to an entangled final state due to its nonseparability. 

\section{Qubit Error mechanisms and quantum measurements}\label{Sec:QubitFidelity}

A critical milestone in achieving large-scale quantum computing is the successful experimental implementation of fault-tolerant quantum logical operations~\cite{KrantzetalQuantumEnginnersGuideAppPhysRev2019,QubitFidelity10.3389/fphy.2022.893507}. When dealing with a two-level system as a qubit, it becomes imperative to execute a series of gate operations while preserving the coherence between the qubit's basis states. In this context, the primary sources of quantum errors leading to qubit decoherence are pure dephasing and spontaneous relaxation from the excited state~\cite{Xue2021,QubitFidelity10.3389/fphy.2022.893507,Book_nielsen_chuang_2010}. Both these mechanisms compromise the operational fidelity of quantum computing processes~\cite{QubitFidelity10.3389/fphy.2022.893507} and reduce the entanglement between qubit states~\cite{HorodeckiRevModPhys81865(2009)QuantumEntanglement}.

To quantify the detrimental effects of spontaneous relaxation and pure dephasing, we numerically solve the Lindblad Master Equation (LME) for the density matrix operator, denoted as $\hat{\rho}=|\psi\rangle\langle\psi|$, or simply the density operator, with $|\psi\rangle$ representing the quantum state of the system. The LME is expressed as follows~\cite{MicrocavitiesBook}:
\begin{equation}
\imath \frac{d\hat{\rho}}{d\tau} = [\hat{\mathcal{H}},\hat{\rho}] + \gamma_{r}\mathcal{L}[\hat{\sigma}_{-}]\hat{\rho} + \gamma_{d}\mathcal{L}[\hat{\sigma}_{z}]\hat{\rho}, \label{eq:MasterEquation}
\end{equation}
where $\hat{\mathcal{H}}$ is either the single-qubit [Eq.~(\ref{eq:H_Sigle_Parametrized})] or two-qubit Hamiltonian [Eq.~(\ref{eq:H_twoqubit})], $ \mathcal{L}[\hat{A}]\hat{\rho} = \hat{A}\hat{\rho}\hat{A}^{\dagger} - \frac{1}{2}\{\hat{A}^{\dagger}\hat{A},\hat{\rho}\}$ and $\hat{\sigma}_{-} =\hat{\sigma}_{x}-\imath \hat{\sigma}_{y}$ is the standard lowering operator. Notice that the operators $\hat{\sigma}_{-}$ and $\hat{\sigma}_{z}$ will be expanded in a $4\times4$ subspace for the two-qubit system, following the definition of Sec.~\ref{TwoQubitGates}. The first term in the right-hand-side of Eq.~\eqref{eq:MasterEquation} accounts for the coherent dynamics of the density operator, while the last two govern the spontaneous relaxation and pure dephasing, with rates $\gamma_{r}$ and $\gamma_{d}$, respectively. As a result of the LME approach, we obtain the final density matrix operator after evolving over a given time interval $\Delta t:0\rightarrow\tau$, i.e., the laser operational time defined in the unitary operator of Eq.~(\ref{eq:UnitaryOp_SingleQubit}) in either the presence or absence of dephasing and relaxation mechanisms.

The phenomenon of spontaneous relaxation corresponds to the transition of the qubit's upper energy state to its lowest state. This type of error primarily impacts the populations of each qubit state, thus manifesting itself in the diagonal elements of the density operator. Conversely, the pure dephasing mechanism is responsible for the degradation of coherent information between the qubit states, resulting in a reduction of the off-diagonal elements of the density operator. In the context of polariton condensates, both relaxation and pure dephasing effects emerge from the interplay between polaritons within the condensate state and their surrounding environment of noncondensed particles~\cite{Askitopoulos_Arxiv2019,Sigurdsson_PRL2022}, such as the scattering of polaritons with the reservoir of hot excitons. 

Decoherence mechanisms due to interaction with the system environment impact all quantum operations within a universal quantum computer~\cite{KrantzetalQuantumEnginnersGuideAppPhysRev2019}, i.e., initialization, quantum gate operation, measurement, and memory, which is realized through the integration of numerous qubits and distinct quantum gates. While tackling the LME for the many-body scenario remains complicated~\cite{QubitFidelity10.3389/fphy.2022.893507,ClerkQuantumNoiseRevModPhys821155(2010),BurkardMultiLevelDecoherencePhysRevB.69.064503(2004)}, it is worth noting that every quantum operation can be decomposed into a series of single and two-qubit gates. This allows us to simplify the analysis of quantum errors in gate operations introduced by decoherence phenomena into a single or two-body problem. 

For a single-qubit gate, cf. Sec.~\ref{SingleQubitGates}, a final unit Bloch vector $\hat{\boldsymbol{u}}=(x,y,z)$ can be directly extracted from the resulting density matrix~\cite{Book_nielsen_chuang_2010}, so that $x = 2\text{Re}(\hat{\rho}_{01})$, $y = 2\text{Im}(\hat{\rho}_{10})$ and $z = \hat{\rho}_{00} - \hat{\rho}_{11}$, where $\hat{\rho}_{ij}$ are the elements $(i,j)$ of the density matrix operator given by Eq.~(\ref{eq:MasterEquation}). In this way, we can compare the initial and final qubit states projected onto the Bloch sphere after its dynamical evolution for a given single qubit gate, both in the presence and absence of quantum errors introduced by the interaction with the surrounding environment and described by Eq.~(\ref{eq:MasterEquation}).

Also, in the context of single-qubit operations, a valuable metric for quantifying the impact of decoherence processes within the trap's environment on the final qubit state, as described by Eq.(\ref{eq:MasterEquation}), is the Von-Neumann entropy~\cite{Book_nielsen_chuang_2010,HorodeckiRevModPhys81865(2009)QuantumEntanglement}. This entropy is defined as:
\begin{equation}
S(\hat{\rho})=-\text{Tr}(\hat{\rho}\log_{2}\hat{\rho}).\label{eq: VN Entropy}   
\end{equation}
When the qubit system is in a perfectly pure state, the Von-Neumann entropy equals zero. The maximum value of the Von-Neumann entropy, denoted as $S(\hat{\rho})=\log_{2}(d)$, is reached when the qubit is in a completely mixed state, with $d$ representing the system's dimensionality~\cite{Book_nielsen_chuang_2010}. For the single-qubit operations defined in Sec.~\ref{SingleQubitGates}, the maximum Von-Neumann entropy is $S_{\text{max}}(\hat{\rho})=\log_{2}(2)=1$.


Another figure-of-merit, due to the presence of quantum error sources, is the so-called fidelity $\in[0,1]$ of a decoherent quantum system compared with its corresponding ideal coherent case ($\gamma_{r}=\gamma_{d}=0$), being computed as follows~\cite{QubitFidelity10.3389/fphy.2022.893507,Kyriienko2016}:
\begin{equation}
 F(\hat{\rho}_{\text{ideal}},\hat{\rho}(t)) = \text{Tr}\left(\sqrt{\sqrt{\hat{\rho}_{\text{ideal}}}\hat{\rho}(t)\sqrt{\hat{\rho}_{\text{ideal}}}}\right).\label{eq:Fidelity} 
\end{equation}
Here, $\hat{\rho}_{\text{ideal}}$ is the density matrix for the pure qubit state and $\hat{\rho}$ is from~\eqref{eq:MasterEquation} in presence of decoherent terms $\gamma_{r}$ and $\gamma_{d}$. 
A perfect fidelity, $F(\hat{\rho}_{\text{ideal}},\hat{\rho}(t))=1$, is only achieved when $\hat{\rho}(t)=\hat{\rho}_{\text{ideal}}$. 

In cases where a two-qubit gate operation results in an entangled state, such as with CNOT and CPHASE operations~\cite{KrantzetalQuantumEnginnersGuideAppPhysRev2019}, it is essential to assess the degree of entanglement in the output state and how it is influenced by dephasing and relaxation mechanisms. This can be achieved by examining the quantum concurrence of the density operator~\cite{WoottersPhysRevLett.80.2245(1998),HorodeckiRevModPhys81865(2009)QuantumEntanglement,Cuevas_SciAdv2018}, defined as follows:
\begin{equation}
\mathcal{C}(\hat{\rho})=\max(0,\lambda_{1}-\lambda_{2}-\lambda_{3}-\lambda_{4}),\label{eq:Concurrence}   
\end{equation}
where $\lambda_{i}$ are the eigenvalues of a matrix $\hat{R}=\sqrt{\sqrt{\hat{\rho}}\tilde{\hat{\rho}}\sqrt{\hat{\rho}}}$ arranged in a decreasing order, with $\tilde{\hat{\rho}}=\hat{\Sigma}\hat{\rho}^{*}\hat{\Sigma}$ being the ``spin-flipped'' density operator and $\hat{\Sigma}=\hat{\sigma}_{y}\otimes\hat{\sigma}_{y}$. The concurrence metric ranges between $0$ and $1$, with a value of $1$ indicating the highest degree of entanglement, meaning a complete mixture of states. Oppositely, a concurrence value of $0$ characterizes separated states, indicating a complete lack of entanglement between the qubits.

\section{Results and Discussion}

We start our analysis by addressing the impacts of both pure dephasing and spontaneous relaxation in the dynamics of a single-qubit Hadamard gate operation, as defined in Eq.~(\ref{eq:Hadamard}). Both the solution of the LME [Eq.~(\ref{eq:MasterEquation})], calculation of fidelity [Eq.~(\ref{eq:Fidelity})], Von-Neumann entropy [Eq.~(\ref{eq: VN Entropy})] and subsequent quantum concurrence [Eq.~(\ref{eq:Concurrence})] were numerically calculated employing the \textit{QuTiP {\textcopyright} package},\textit{ a Quantum Toolbox in Python, version 4.7.1}~\cite{QuTiP2JOHANSSON20131234,QuTip1JOHANSSON20121760}.

\begin{figure}[t]
\centerline{\includegraphics[width=0.51\textwidth,keepaspectratio]{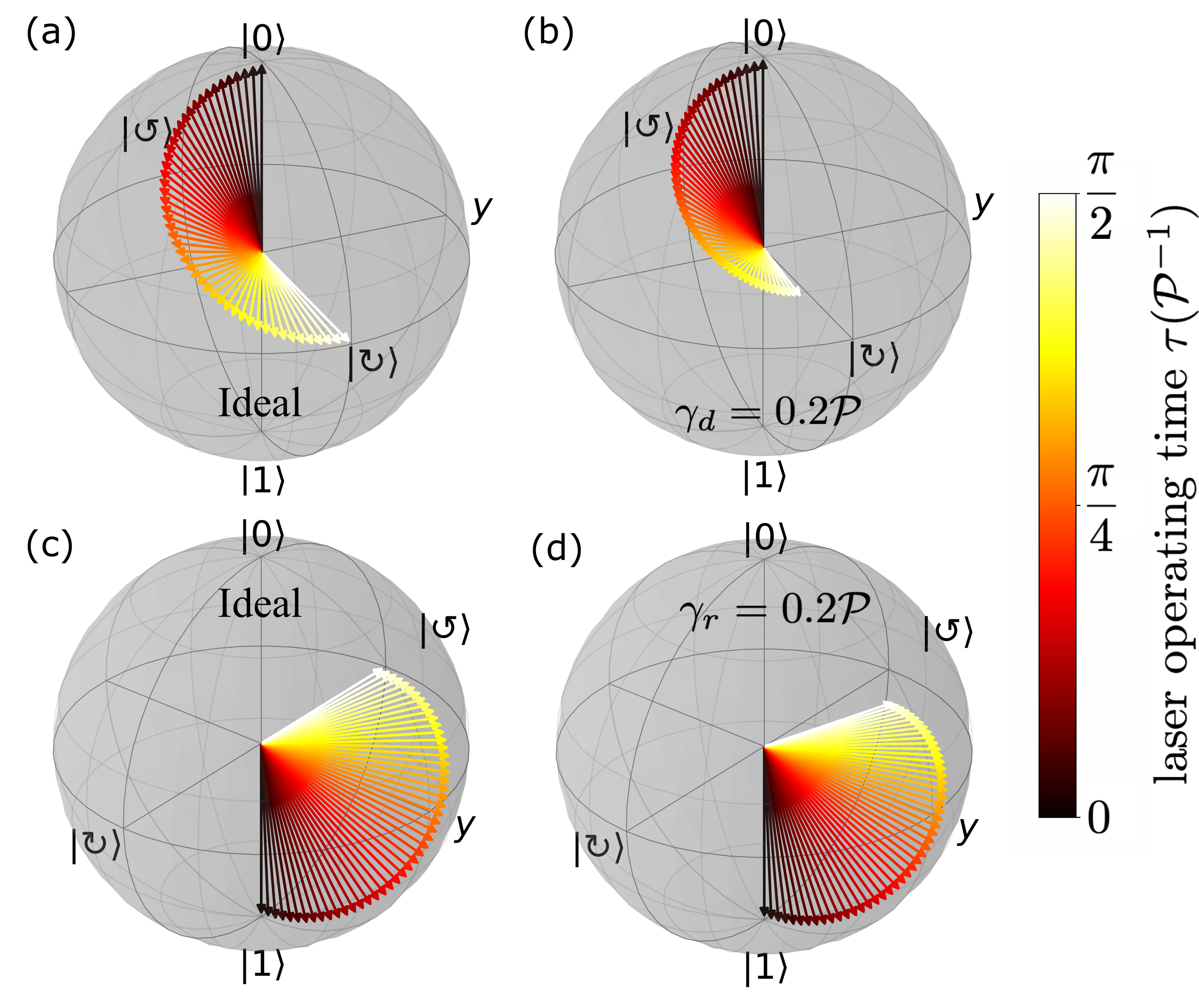}}
\caption{\label{Fig:BlochSphereHadamard} Evolution of the initial single-qubit state projected as a unit vector onto the Bloch sphere, under the application of the Hadamard gate defined in Eq.~(\ref{eq:Hadamard}), considering that the auxiliary laser beam was applied for a period of time $\tau = \frac{\pi}{2\mathcal{P}}$. Panels (a) and (b) depict the dynamics of the initial qubit state $|0\rangle \equiv |p_x \rangle $ for a Hadamard gate, in the absence and presence of pure dephasing, respectively. The same is shown in panels (c) and (d), but with $|1\rangle \equiv |p_y \rangle$ as the single-qubit initial state, now considering the presence and absence of spontaneous relaxation rate, respectively. The lateral color bar indicates the corresponding timescale for each Bloch vector, from the initial time $\tau=0$ to the final time $\frac{\pi}{2\mathcal{P}}$.}
\end{figure}

As stated by the unitary operator of Eq.~(\ref{eq:Hadamard}), to perform a Hadamard gate, the phase of the auxiliary laser beam $\theta =0$, with an operating time $\tau = \frac{\pi}{2\mathcal{P}}$ and $\phi = \frac{\pi}{4}$. This last condition implies a non-zero laser beam-qubit detuning, once $\cos{\phi}=\Delta \varepsilon / 2\mathcal{P}$. As it can be noticed, the parameters are in units of $\mathcal{P}$, which may vary from one experimental system to another depending on the parameters of the microcavity, photon-exciton detuning, and exciton oscillator strength.

Fig.~\ref{Fig:BlochSphereHadamard}(a) shows the evolution of the initial qubit state $|\psi\rangle=|0\rangle$, or equivalently $|p_{x}\rangle$, projected onto a Bloch sphere when submitted to a Hadamard operation, as defined in Eq.~(\ref{eq:Hadamard}), in the absence of any quantum error source ($\gamma_{r} = \gamma_{d} = 0$). It can be noticed that the action of the Hadamard gate in the qubit basis state $|0\rangle$ creates an equal symmetric superposition of the two-basis states $(|0\rangle + |1\rangle)/\sqrt{2} \equiv |\circlearrowright \rangle$, despite the presence of a phase of $-\frac{\pi}{2}$, as shown in Eq.~(\ref{eq:Hadamard_to_|zero>}). The resulting qubit state is represented by a unit vector ($|\hat{\boldsymbol{u}}|=1$) residing within the equatorial plane of the Bloch sphere, thereby defining a pure state.

The contrasting case of Fig.~\ref{Fig:BlochSphereHadamard}(a) is depicted in the corresponding panel (b), where the pure dephasing mechanism as an error source is considered, specifically for $\gamma_{d}=0.2\mathcal{P}$. A direct comparison between Figs.~\ref{Fig:BlochSphereHadamard}(a) and (b) reveals the effects of pure dephasing on the final qubit state following the application of the Hadamard gate. Upon the completion of the laser beam operation within the time interval $\tau$, the final state diverges from the vortex mode $|\circlearrowright\rangle$ and is now characterized by a vector state $\hat{\boldsymbol{u}}$ located within the Bloch sphere, with $|\hat{\boldsymbol{u}}|\approx 0.70$. This reduction in the norm of the Bloch vector indicates that the final qubit state is in a mixed state. This mixing is a consequence of the presence of the pure dephasing mechanism, which introduces non-zero values in the off-diagonal elements of the density operator.

\begin{figure}[t]
\centerline{\includegraphics[width=0.51\textwidth,keepaspectratio]{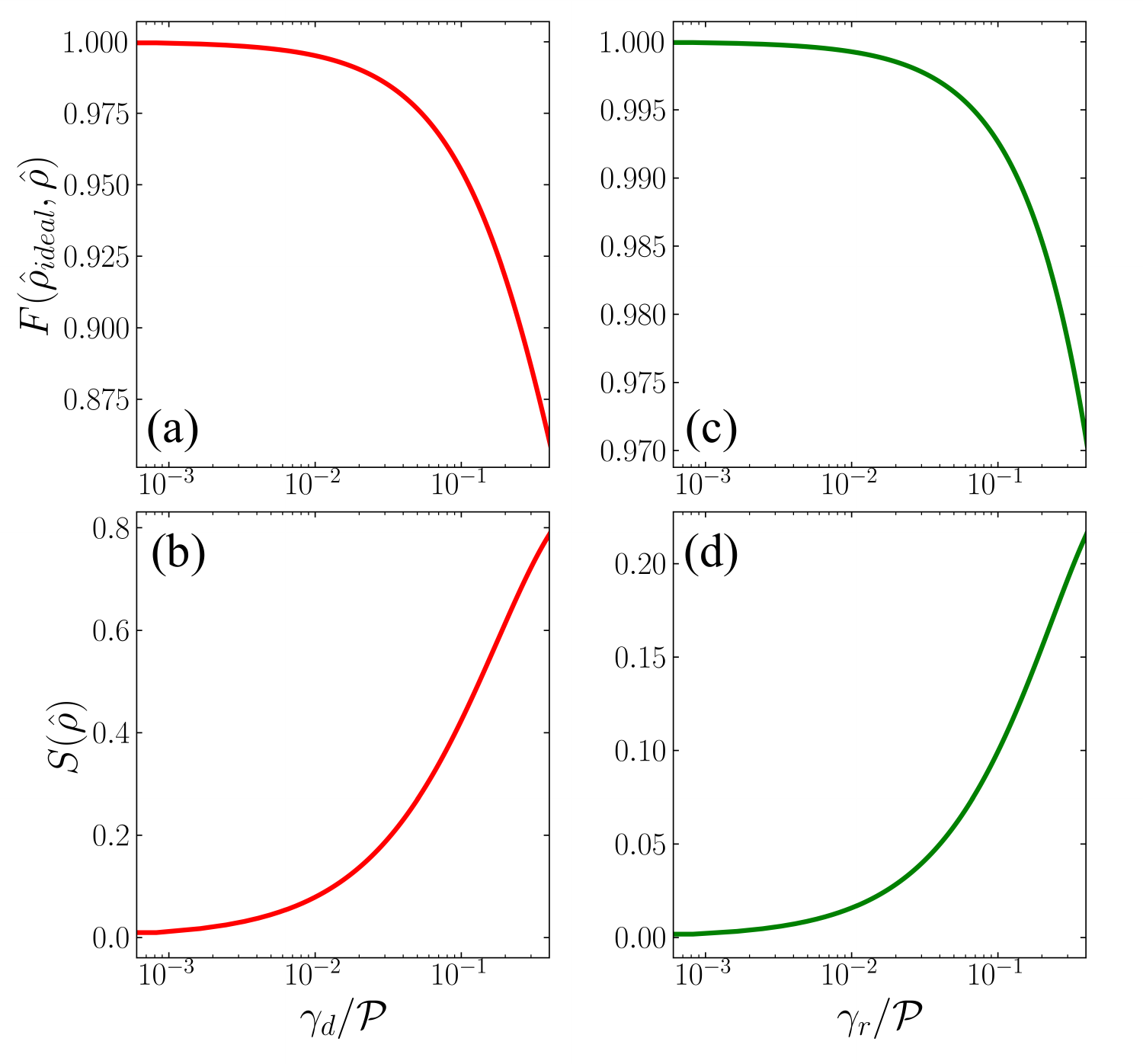}}
\caption{\label{Fig:FidelityHadamardSingleQubit} Single-qubit state fidelity [Eq.~(\ref{eq:Fidelity})] and Von-Neumann entropy for a Hadamard gate operation [Eq.~(\ref{eq:Hadamard})] as a function of either pure dephasing [panels (a)-(b)] or spontaneous relaxation [panels (c)-(d)] rates $\gamma_d$ and $\gamma_r$, respectively, considering the qubit initial state $|1\rangle$, with $x$-axis on the logarithmic scale. Both the rates vary between $0$ and $0.4\mathcal{P}$.}
\end{figure}

\begin{figure}[t]
\centerline{\includegraphics[width=0.51\textwidth,keepaspectratio]{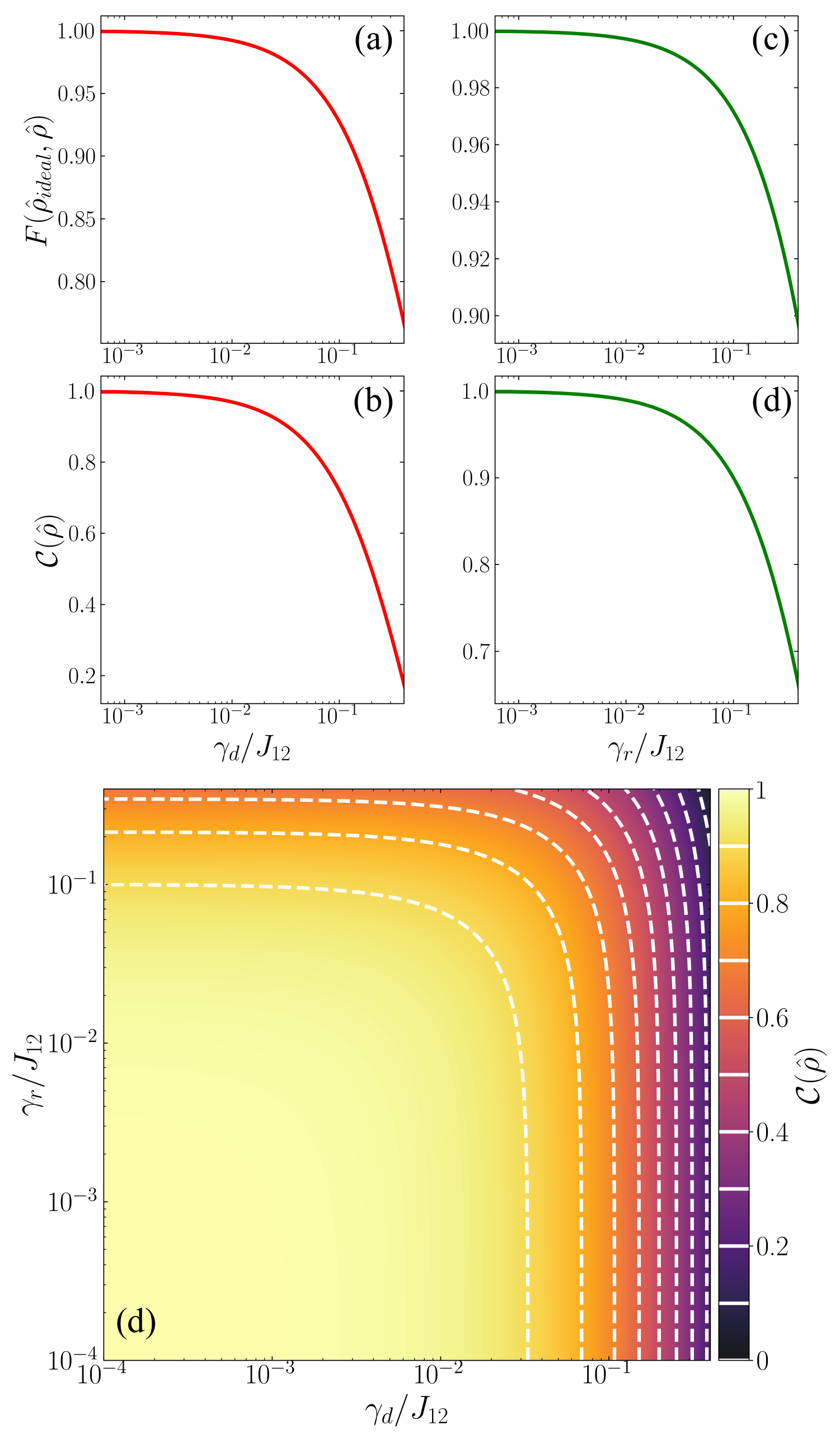}}
\caption{\label{Fig:Fidelity_and_Concurrence_CPHASE_gate}Two-qubit state fidelity [Eq.~(\ref{eq:Fidelity})] and quantum concurrence [Eq.~(\ref{eq:Concurrence})] for a CPHASE gate operation as a function of pure dephasing [panels (a)-(b)] or spontaneous relaxation [panels (c)-(d)] rates $\gamma_d$ and $\gamma_r$, respectively, with the \textit{x}-axis on the logarithmic scale. The two-qubit initial state is set up as $|\psi_{0}\rangle =|\circlearrowright\rangle_{\text{C}}|\circlearrowleft\rangle_{\text{T}}$, resulting in an entangled final state, as shown in Eq.~(\ref{eq:CPHASE_final}). Panel (e) maps the behavior of the concurrence as a function of both $\gamma_d$ and $\gamma_r$ for the same CPHASE gate applied in $|\psi_{0}\rangle$.}
\end{figure}

The application of the Hadamard gate in the single-qubit state $|1\rangle\equiv|p_{y}\rangle$ is depicted in Fig.~\ref{Fig:BlochSphereHadamard}(c), in which we can verify that the final state is $|\circlearrowleft\rangle$, i,e., an anticlockwise current mode, as expected for the Hadamard operation. By considering now the pure relaxation of the qubit initial state, with $\gamma_{r}=0.2\mathcal{P}$, Fig.~\ref{Fig:BlochSphereHadamard}(d) points out that the qubit final state vector after $\tau$ is not in the same position of $|\circlearrowleft\rangle$, also with a small reduction of its vector state ($|\hat{\boldsymbol{u}}|\approx 0.96$), also indicating some degree of mixture. 

To enhance our comprehension of the impacts stemming from pure relaxation and dephasing, as depicted in the Bloch sphere (Fig.~\ref{Fig:BlochSphereHadamard}) for the Hadamard gate, in Fig.~\ref{Fig:FidelityHadamardSingleQubit} we delve we explore the fidelity and Von-Neumann entropy for the resulting single-qubit state, as outlined in Eqs.(\ref{eq:Fidelity}) and~(\ref{eq: VN Entropy}). In Fig.~\ref{Fig:FidelityHadamardSingleQubit}(a), we observe a notable exponential reduction in fidelity as the pure dephasing rate $\gamma_{d}$ increases. Simultaneously, the Von-Neumann entropy exhibits a corresponding increase, beginning at the value of $0$ when $\gamma_{d}=0$, indicating a completely pure state. As $\gamma_{d}$ progresses to $0.4\mathcal{P}$, the Von-Neumann entropy approaches $1$, indicating the formation of an almost entirely mixed state. This strong enhancement of the Von-Neumann entropy is in agreement with the pure dephasing mechanism, which mixes the basis states of the qubit.

Figs.~\ref{Fig:FidelityHadamardSingleQubit}(c) and (d) exhibit a similar pattern to their counterparts in panels (a) and (b) regarding fidelity and Von-Neumann entropy, but as functions of the pure spontaneous relaxation rate $\gamma_r$. It is worth noting that while the impact of pure relaxation on the final qubit state appears to be less severe when compared to pure dephasing, it is crucial to emphasize that in the case of pure relaxation, the qubit exchanges energy with its surrounding environment, leading to a complete loss of information, characterizing an irreversible process~\cite{QubitFidelity10.3389/fphy.2022.893507}.

Our analysis now shifts towards the operation of a two-qubit gate, specifically examining the impact of pure dephasing and spontaneous relaxation in the CPHASE operation, as defined by Eq.~(\ref{eq:CPHASE}). As previously mentioned, the CPHASE gate falls into the category of \textit{entangling gates}, as its application leads to a final state that cannot be decomposed into individual single-qubit states. This characteristic makes the CPHASE gate particularly suitable for investigating quantum concurrence [see Eq.~(\ref{eq:Concurrence})].

In Fig.~\ref{Fig:Fidelity_and_Concurrence_CPHASE_gate} we explore how the two-qubit state fidelity and concurrence changes under the application of a CPHASE gate when both the spontaneous relaxation and dephasing mechanisms are accounted equally in both neighboring traps. The two-qubit initial state is prepared in the same state employed in Eq.~(\ref{eq:CPHASE_final}), which leads to an entangled state after the application of the CPHASE gate as defined in Eq.~(\ref{eq:CPHASE}). Fig.~\ref{Fig:Fidelity_and_Concurrence_CPHASE_gate}(a) and (b) show both the fidelity of the final qubit state and quantum concurrence as a function of pure dephasing. In addition to the expected reduction of the fidelity in Fig.~\ref{Fig:Fidelity_and_Concurrence_CPHASE_gate}(a) when compared to the ideal case ($\gamma_{d}=\gamma_{r}=0$), in Fig.~\ref{Fig:Fidelity_and_Concurrence_CPHASE_gate}(b) we observe a progressive degradation of the entangled output state, indicated by the exponential decline in concurrence. It's worth noting that when $\gamma_{d}=0$, the entanglement between the qubit states reaches its maximum, with $\mathcal{C}(\hat{\rho})=1$. However, as the pure dephasing rate increases, the off-diagonal (coherence) elements of the density operators undergo significant reduction, leading to a nearly complete loss of entanglement in the system. This is evident in the upper limit case of $\gamma_{d}=0.4J_{12}$, where $\mathcal{C}(\hat{\rho})\approx 0.16$, corresponding to a final state fidelity of approximately 0.76, as depicted in Fig.~\ref{Fig:Fidelity_and_Concurrence_CPHASE_gate}(a).

Fig.~\ref{Fig:Fidelity_and_Concurrence_CPHASE_gate}(c) and (d) also show the fidelity and concurrence of the final entangled two-qubit state, but for the case in which the CPHASE gate operation is subjected to a spontaneous relaxation mechanism. We can verify the reduction of both quantities as the spontaneous relaxation rate $\gamma_{r}$ increases. As the spontaneous relaxation rate $\gamma_{r}$ increases, both fidelity and concurrence noticeably decrease. However, it is noteworthy that for the higher value of $\gamma_{r}$ considered, both the output state fidelity and concurrence surpass those observed in the case of pure dephasing, as depicted in Figs.~\ref{Fig:Fidelity_and_Concurrence_CPHASE_gate}(a)-(b). Specifically, for $\gamma_{r}=0.4J_{12}$, the fidelity reaches approximately 0.90, while $\mathcal{C}(\hat{\rho})\approx 0.66$. This significant difference between the effects of pure dephasing [Figs.\ref{Fig:Fidelity_and_Concurrence_CPHASE_gate}(a)-(b)] and spontaneous relaxation [Figs.\ref{Fig:Fidelity_and_Concurrence_CPHASE_gate}(c)-(d)] arises from the distinct impact that each mechanism has on the density operator. Pure dephasing primarily reduces the off-diagonal elements of the density operator, which encode information about quantum coherence between distinct qubits. In contrast, spontaneous relaxation predominantly affects the diagonal elements (populations) of the density operator, leaving the off-diagonal elements relatively unchanged.

To provide a comprehensive overview of the concurrence behavior in the presence of both pure dephasing and spontaneous relaxation during the application of the CPHASE gate, we present a concurrence colormap in Fig.~\ref{Fig:Fidelity_and_Concurrence_CPHASE_gate}(e). This panel reveals an optimal range for both rates where entanglement between the two qubits is almost entirely preserved, specifically $0.90\leq\mathcal{C}(\hat{\rho})\leq 1$, for $\gamma_{d}\leq 0.03J_{12}$ and $\gamma_{r}\leq 0.1J_{12}$. This region is enclosed by the first white dashed line in Fig.~\ref{Fig:Fidelity_and_Concurrence_CPHASE_gate}(e).

\section{Conclusions and Outlook}

In this work, we explore the possibility of employing elliptically trapped polariton condensates as quantum logic gates. The single-qubit basis states $|0\rangle$ and $|1\rangle$ in each condensate are defined as being the $|p_x\rangle$ and $|p_y\rangle$ states, respectively. These states are associated with the spatial dipolar distributions of the polariton density along the orthogonal axes of the trap, while their energy splitting can be adjusted through the geometrical ellipticity of the trap via SLMs. Distinct linear combinations between these $p$-states describe polariton condensates with integer OAM, which carry clockwise and anticlockwise superfluid current modes.

By introducing an auxiliary laser beam in each trap, we demonstrate the feasibility of implementing a versatile set of universal single-qubit operations, including Pauli and Hadamard gates. Furthermore, by exploring different types of interaction between the two traps, we unveil the potential for executing two-qubit gate operations such as CPHASE and CNOT gates. These two-qubit gates fall into the crucial category of entangling gates, which are fundamental for quantum computation.

We also investigate how quantum error sources, common in polariton condensates such as pure dephasing and spontaneous relaxation from the qubit's excited state, impact the final state of the proposed two-qubit system, particularly for Hadamard and CPHASE gate operations. To address this, we numerically compute the corresponding density operator via the Master Equation approach. Subsequently, we assess key performance metrics, including the fidelity of the final state, Von-Neumann entropy, and quantum concurrence.

In the context of local quantum computing operations, it is important to ensure that any proposal of practical implementation of a quantum computer must satisfy the DiVincenzo's criteria~\cite{DiVincenzo2000}, which are, \textit{ipsis litteris}, the following: \textit{(i) A scalable physical system with well-characterized qubits; (ii) The ability to initialize the state of the qubits to a simple fiducial state, such as $|000\ldots \rangle$; (iii) Long relevant decoherence times, much longer than the gate operation time; (iv) A universal set of quantum gates and (v) A qubit-specific measurement capability.} Below, we briefly discuss each of these criteria within our proposal, with the exception of criterion \textit{(iv)}, which constitutes the main finding of the current work. This concise discussion aims to outline potential avenues for realizing the proposed optically trapped polariton condensate qubits.

The criterion \textit{(i)} means, at first, that a quantum computer must contain several quantum bits, which in the context of our system, implies an optical device with a collection of coupled traps of polariton condensates. Furthermore, each qubit (trap) must have its physical parameters well characterized, as the qubit Hamiltonian, the presence of couplings between distinct qubits and interaction with external fields which can be used for initializing the qubit state and performing quantum logic operations. Considering both these features, we drawn attention for the work of Alyatkin \textit{et al.}~\cite{alyatkin2022alloptical} which have experimentally explored a triangular lattice of all-optical driven trapped polariton condensates with integer OAM $l=\pm 1$ carrying vortex ($l=+1$) and antivortex ($l=-1$) states with a Ising-type interaction between them. This system could be extended to encompass several traps of polariton condensates, effectively creating a multiple-qubit setup, with optical tunability for each individual component, as described by Eq.~(\ref{eq:single_qubit}). Moreover, the Ising-type interaction between trapped condensates at distinct lattice sites favors the implementation of a CPHASE gate, cf. Eq.~(\ref{eq:CPHASE}).

The capability to set the initial qubit states, as indicated by criterion \textit{(ii)}, arises from the obvious prerequisite that input states must be precisely defined before implementing any quantum computing operation. Furthermore, the specific need to initialize the system in the global ground state $|000\ldots \rangle$, equivalent to $|p_{x}p_{x}p_{x}\ldots \rangle$ within our proposal, is required by an effective quantum error correction implementation, which demands a constant and pristine source of qubits in a low-entropy state~\cite{DiVincenzo2000}. In the context of the device sketched in Fig.~\ref{Fig:Device}, the direction of a polariton vortex can be controlled by means of an ultra-short (120 fs) off-resonant optical control pulse~\cite{Assmann1}, for instance, in which the final direction of the vortex depends on the power and duration of this pulse. Trapped polariton condensates also can be effectively controlled by external rotating potentials, as evidenced by recent experimental findings~\cite{Gnusov_SciAdv2023,delValle_NanoLetters_2023} and theoretically described by Yulin \textit{et al.}~\cite{yulin2023vorticity}. Furthermore, we draw attention to the possibility of experimental tuning of the coupling between neighboring polariton traps within a two-dimensional network through all-optical methods, as detailed in Ref.~\onlinecite{Alyatkin_PhysRevLett124207402(2020)}. This result holds significant implications for the realization of two-qubit quantum gates, as detailed in Section~\ref{TwoQubitGates}.  


Regarding point \textit{(iii)}, the decoherence time plays a pivotal role in the dynamics of any quantum system coupled to its surrounding environment, once it indicates for how long the quantum behavior is preserved before succumbing to the classical one. This means that the decoherence time $\tau_{\text{coh}}$ for a quantum gate operation should be long enough to ensure that the quantum coherence of the system is preserved, which implies that it should be longer than the time to perform a quantum operation. In our analysis of single and double-qubit gates, the ``clock-time'' for each corresponding operation is given by $\tau$. Specifically, for a single operation $\tau \sim 10^{-5},10^{-4}\tau_{\text{coh}}$~\cite{DiVincenzo2000}, in order to preserve the quantum nature of the computing process and also ensure the performance of error correction mechanisms. In the scenario of polariton condensates, the decoherence time can reach several ns~\cite{Caputo_NatMat2018,Askitopoulos_Arxiv2019,Sigurdsson_PRL2022}, which is related to the time-dependent fluctuations of the overall phase of the condensate wave function. However, the spatial coherence of an optically trapped polariton condensate driven by a CW pump has been reported as practically uniform~\cite{Topfer_PRB2020,Sigurdsson_PRL2022,Sedov2021}. Notably, in an experiment conducted by Sedov \textit{et al.}~\cite{Sedov2021}, no traces of any loss due to spatial decoherence were detected throughout the entire duration of the optical experiment, suggesting a decoherence time on the order of milliseconds ($\tau_{\text{coh}} \sim \text{ms}$).

Finally, as addressed by criterion \textit{(v)}, the outcomes of a quantum computing process must be read out. In other words, the probability outcomes associated with states of the system encoded by the density operator must be experimentally accessed. In the framework of our proposal, the information of the final two-qubit states can be detected by implementing a protocol inspired by the work of Mair \textit{et al.}~\cite{ZeilingerNature412_313-316(2001)}, originally devised for characterizing photons with entangled OAM. As a first step, the light emitted by each of the traps of polariton condensates would pass through an SLM, where a predefined function is applied to manipulate the spatial phase distribution of the light. These preset functions would correspond to either the circular current states or the dipolar \textit{p}-states, oriented along the main axis and short axis of the corresponding elliptical trap. Subsequently, once the light has passed through the SLM, it is directed into a single-mode optical fiber. This optical fiber operates as a filter, allowing only the Gaussian mode to propagate while suppressing all other modes. At the termination point of each optical fiber, a photodetector is positioned and calibrated to generate a binary signal. This binary signal is designed to indicate a value of 1 when the intensity of the Gaussian mode surpasses a predefined critical threshold, and it registers 0 when the intensity falls below this threshold. Then, a coincidence counter compares the signal coming from both photodetectors. If both record the value of 1, the counter computes the value of 1; conversely, it records 0. By considering the four possible preset functions for each of the two SLMs, we obtain 16 independent measurements of the states of two qubits projected along the $z$ or $x$ axes of the corresponding Bloch spheres. By accumulating statistics over thousands of measurements, a high-fidelity measurement matrix of the system is obtained~\cite{PlantenbergNature2007,Book_nielsen_chuang_2010}. This matrix can then be converted into the corresponding density operator, which encodes the probabilities of the system states.


\begin{acknowledgments}
L.S.R. and I.A.S. acknowledge the support from the Icelandic Research Fund (Rann\'{i}s), grant No. 163082-051 and Project Hybrid Polaritonics. A.K. acknowledges the support of the Russian Foundation for Basic Research (Grant No. 19-52-12032). We acknowledge Aleksey Fedorov, Helgi Sigur{\dh}sson, and Boris Altshuler for fruitful discussions. We also acknowledge Roman Cherbunin for designing the read-out scheme of a two-qubit polariton gate. 
\end{acknowledgments}

\bibliography{Refs}

\begin{thebibliography}{73}%
\makeatletter
\providecommand \@ifxundefined [1]{%
 \@ifx{#1\undefined}
}%
\providecommand \@ifnum [1]{%
 \ifnum #1\expandafter \@firstoftwo
 \else \expandafter \@secondoftwo
 \fi
}%
\providecommand \@ifx [1]{%
 \ifx #1\expandafter \@firstoftwo
 \else \expandafter \@secondoftwo
 \fi
}%
\providecommand \natexlab [1]{#1}%
\providecommand \enquote  [1]{``#1''}%
\providecommand \bibnamefont  [1]{#1}%
\providecommand \bibfnamefont [1]{#1}%
\providecommand \citenamefont [1]{#1}%
\providecommand \href@noop [0]{\@secondoftwo}%
\providecommand \href [0]{\begingroup \@sanitize@url \@href}%
\providecommand \@href[1]{\@@startlink{#1}\@@href}%
\providecommand \@@href[1]{\endgroup#1\@@endlink}%
\providecommand \@sanitize@url [0]{\catcode `\\12\catcode `\$12\catcode
  `\&12\catcode `\#12\catcode `\^12\catcode `\_12\catcode `\%12\relax}%
\providecommand \@@startlink[1]{}%
\providecommand \@@endlink[0]{}%
\providecommand \url  [0]{\begingroup\@sanitize@url \@url }%
\providecommand \@url [1]{\endgroup\@href {#1}{\urlprefix }}%
\providecommand \urlprefix  [0]{URL }%
\providecommand \Eprint [0]{\href }%
\providecommand \doibase [0]{https://doi.org/}%
\providecommand \selectlanguage [0]{\@gobble}%
\providecommand \bibinfo  [0]{\@secondoftwo}%
\providecommand \bibfield  [0]{\@secondoftwo}%
\providecommand \translation [1]{[#1]}%
\providecommand \BibitemOpen [0]{}%
\providecommand \bibitemStop [0]{}%
\providecommand \bibitemNoStop [0]{.\EOS\space}%
\providecommand \EOS [0]{\spacefactor3000\relax}%
\providecommand \BibitemShut  [1]{\csname bibitem#1\endcsname}%
\let\auto@bib@innerbib\@empty
\bibitem [{\citenamefont {Kavokin}\ \emph {et~al.}(2022)\citenamefont
  {Kavokin}, \citenamefont {Liew}, \citenamefont {Schneider}, \citenamefont
  {Lagoudakis}, \citenamefont {Klembt},\ and\ \citenamefont
  {Hoefling}}]{KavokinNatRevPhys2022}%
  \BibitemOpen
  \bibfield  {author} {\bibinfo {author} {\bibfnamefont {A.}~\bibnamefont
  {Kavokin}}, \bibinfo {author} {\bibfnamefont {T.~C.~H.}\ \bibnamefont
  {Liew}}, \bibinfo {author} {\bibfnamefont {C.}~\bibnamefont {Schneider}},
  \bibinfo {author} {\bibfnamefont {P.~G.}\ \bibnamefont {Lagoudakis}},
  \bibinfo {author} {\bibfnamefont {S.}~\bibnamefont {Klembt}},\ and\ \bibinfo
  {author} {\bibfnamefont {S.}~\bibnamefont {Hoefling}},\ }\bibfield  {title}
  {\bibinfo {title} {Polariton condensates for classical and quantum
  computing},\ }\href {https://doi.org/10.1038/s42254-022-00447-1} {\bibfield
  {journal} {\bibinfo  {journal} {Nature Reviews Physics}\ }\textbf {\bibinfo
  {volume} {4}},\ \bibinfo {pages} {435} (\bibinfo {year} {2022})}\BibitemShut
  {NoStop}%
\bibitem [{\citenamefont {Liew}(2023)}]{Liew_OptMatExp2023}%
  \BibitemOpen
  \bibfield  {author} {\bibinfo {author} {\bibfnamefont {T.~C.~H.}\
  \bibnamefont {Liew}},\ }\bibfield  {title} {\bibinfo {title} {The future of
  quantum in polariton systems: opinion},\ }\href
  {https://doi.org/10.1364/OME.492503} {\bibfield  {journal} {\bibinfo
  {journal} {Opt. Mater. Express}\ }\textbf {\bibinfo {volume} {13}},\ \bibinfo
  {pages} {1938} (\bibinfo {year} {2023})}\BibitemShut {NoStop}%
\bibitem [{\citenamefont {Sanvitto}\ and\ \citenamefont
  {K{\'e}na-Cohen}(2016)}]{Sanvitto_NatMat2016}%
  \BibitemOpen
  \bibfield  {author} {\bibinfo {author} {\bibfnamefont {D.}~\bibnamefont
  {Sanvitto}}\ and\ \bibinfo {author} {\bibfnamefont {S.}~\bibnamefont
  {K{\'e}na-Cohen}},\ }\bibfield  {title} {\bibinfo {title} {The road towards
  polaritonic devices},\ }\href {https://doi.org/10.1038/nmat4668} {\bibfield
  {journal} {\bibinfo  {journal} {Nature Materials}\ }\textbf {\bibinfo
  {volume} {15}},\ \bibinfo {pages} {1061} (\bibinfo {year}
  {2016})}\BibitemShut {NoStop}%
\bibitem [{\citenamefont {Keeling}\ and\ \citenamefont
  {K\'{e}na-Cohen}(2020)}]{Keeling_AnnRev2020}%
  \BibitemOpen
  \bibfield  {author} {\bibinfo {author} {\bibfnamefont {J.}~\bibnamefont
  {Keeling}}\ and\ \bibinfo {author} {\bibfnamefont {S.}~\bibnamefont
  {K\'{e}na-Cohen}},\ }\bibfield  {title} {\bibinfo {title} {Bose–einstein
  condensation of exciton-polaritons in organic microcavities},\ }\href
  {https://doi.org/10.1146/annurev-physchem-010920-102509} {\bibfield
  {journal} {\bibinfo  {journal} {Annual Review of Physical Chemistry}\
  }\textbf {\bibinfo {volume} {71}},\ \bibinfo {pages} {435} (\bibinfo {year}
  {2020})}\BibitemShut {NoStop}%
\bibitem [{\citenamefont {Schneider}\ \emph {et~al.}(2016)\citenamefont
  {Schneider}, \citenamefont {Winkler}, \citenamefont {Fraser}, \citenamefont
  {Kamp}, \citenamefont {Yamamoto}, \citenamefont {Ostrovskaya},\ and\
  \citenamefont {H{\"o}fling}}]{Schneider2017RPP}%
  \BibitemOpen
  \bibfield  {author} {\bibinfo {author} {\bibfnamefont {C.}~\bibnamefont
  {Schneider}}, \bibinfo {author} {\bibfnamefont {K.}~\bibnamefont {Winkler}},
  \bibinfo {author} {\bibfnamefont {M.~D.}\ \bibnamefont {Fraser}}, \bibinfo
  {author} {\bibfnamefont {M.}~\bibnamefont {Kamp}}, \bibinfo {author}
  {\bibfnamefont {Y.}~\bibnamefont {Yamamoto}}, \bibinfo {author}
  {\bibfnamefont {E.~A.}\ \bibnamefont {Ostrovskaya}},\ and\ \bibinfo {author}
  {\bibfnamefont {S.}~\bibnamefont {H{\"o}fling}},\ }\bibfield  {title}
  {\bibinfo {title} {Exciton-polariton trapping and potential landscape
  engineering},\ }\href@noop {} {\bibfield  {journal} {\bibinfo  {journal}
  {Rep. Prog. Phys.}\ }\textbf {\bibinfo {volume} {80}},\ \bibinfo {pages}
  {016503} (\bibinfo {year} {2016})}\BibitemShut {NoStop}%
\bibitem [{\citenamefont {T\"{o}pfer}\ \emph {et~al.}(2021)\citenamefont
  {T\"{o}pfer}, \citenamefont {Chatzopoulos}, \citenamefont {Sigurdsson},
  \citenamefont {Cookson}, \citenamefont {Rubo},\ and\ \citenamefont
  {Lagoudakis}}]{Topfer_Optica2021}%
  \BibitemOpen
  \bibfield  {author} {\bibinfo {author} {\bibfnamefont {J.~D.}\ \bibnamefont
  {T\"{o}pfer}}, \bibinfo {author} {\bibfnamefont {I.}~\bibnamefont
  {Chatzopoulos}}, \bibinfo {author} {\bibfnamefont {H.}~\bibnamefont
  {Sigurdsson}}, \bibinfo {author} {\bibfnamefont {T.}~\bibnamefont {Cookson}},
  \bibinfo {author} {\bibfnamefont {Y.~G.}\ \bibnamefont {Rubo}},\ and\
  \bibinfo {author} {\bibfnamefont {P.~G.}\ \bibnamefont {Lagoudakis}},\
  }\bibfield  {title} {\bibinfo {title} {Engineering spatial coherence in
  lattices of polariton condensates},\ }\href
  {https://doi.org/10.1364/OPTICA.409976} {\bibfield  {journal} {\bibinfo
  {journal} {Optica}\ }\textbf {\bibinfo {volume} {8}},\ \bibinfo {pages} {106}
  (\bibinfo {year} {2021})}\BibitemShut {NoStop}%
\bibitem [{\citenamefont {Kuriakose}\ \emph {et~al.}(2022)\citenamefont
  {Kuriakose}, \citenamefont {Walker}, \citenamefont {Dowling}, \citenamefont
  {Kyriienko}, \citenamefont {Shelykh}, \citenamefont {St-Jean}, \citenamefont
  {Zambon}, \citenamefont {Lema{\^i}tre}, \citenamefont {Sagnes}, \citenamefont
  {Legratiet}, \citenamefont {Harouri}, \citenamefont {Ravets}, \citenamefont
  {Skolnick}, \citenamefont {Amo}, \citenamefont {Bloch},\ and\ \citenamefont
  {Krizhanovskii}}]{Kuriakose_NatPhot2022}%
  \BibitemOpen
  \bibfield  {author} {\bibinfo {author} {\bibfnamefont {T.}~\bibnamefont
  {Kuriakose}}, \bibinfo {author} {\bibfnamefont {P.~M.}\ \bibnamefont
  {Walker}}, \bibinfo {author} {\bibfnamefont {T.}~\bibnamefont {Dowling}},
  \bibinfo {author} {\bibfnamefont {O.}~\bibnamefont {Kyriienko}}, \bibinfo
  {author} {\bibfnamefont {I.~A.}\ \bibnamefont {Shelykh}}, \bibinfo {author}
  {\bibfnamefont {P.}~\bibnamefont {St-Jean}}, \bibinfo {author} {\bibfnamefont
  {N.~C.}\ \bibnamefont {Zambon}}, \bibinfo {author} {\bibfnamefont
  {A.}~\bibnamefont {Lema{\^i}tre}}, \bibinfo {author} {\bibfnamefont
  {I.}~\bibnamefont {Sagnes}}, \bibinfo {author} {\bibfnamefont
  {L.}~\bibnamefont {Legratiet}}, \bibinfo {author} {\bibfnamefont
  {A.}~\bibnamefont {Harouri}}, \bibinfo {author} {\bibfnamefont
  {S.}~\bibnamefont {Ravets}}, \bibinfo {author} {\bibfnamefont {M.~S.}\
  \bibnamefont {Skolnick}}, \bibinfo {author} {\bibfnamefont {A.}~\bibnamefont
  {Amo}}, \bibinfo {author} {\bibfnamefont {J.}~\bibnamefont {Bloch}},\ and\
  \bibinfo {author} {\bibfnamefont {D.~N.}\ \bibnamefont {Krizhanovskii}},\
  }\bibfield  {title} {\bibinfo {title} {Few-photon all-optical phase rotation
  in a quantum-well micropillar cavity},\ }\href
  {https://doi.org/10.1038/s41566-022-01019-6} {\bibfield  {journal} {\bibinfo
  {journal} {Nature Photonics}\ }\textbf {\bibinfo {volume} {16}},\ \bibinfo
  {pages} {566} (\bibinfo {year} {2022})}\BibitemShut {NoStop}%
\bibitem [{\citenamefont {Boulier}\ \emph {et~al.}(2014)\citenamefont
  {Boulier}, \citenamefont {Bamba}, \citenamefont {Amo}, \citenamefont
  {Adrados}, \citenamefont {Lemaitre}, \citenamefont {Galopin}, \citenamefont
  {Sagnes}, \citenamefont {Bloch}, \citenamefont {Ciuti}, \citenamefont
  {Giacobino},\ and\ \citenamefont {Bramati}}]{Boulier_NatComm2014}%
  \BibitemOpen
  \bibfield  {author} {\bibinfo {author} {\bibfnamefont {T.}~\bibnamefont
  {Boulier}}, \bibinfo {author} {\bibfnamefont {M.}~\bibnamefont {Bamba}},
  \bibinfo {author} {\bibfnamefont {A.}~\bibnamefont {Amo}}, \bibinfo {author}
  {\bibfnamefont {C.}~\bibnamefont {Adrados}}, \bibinfo {author} {\bibfnamefont
  {A.}~\bibnamefont {Lemaitre}}, \bibinfo {author} {\bibfnamefont
  {E.}~\bibnamefont {Galopin}}, \bibinfo {author} {\bibfnamefont
  {I.}~\bibnamefont {Sagnes}}, \bibinfo {author} {\bibfnamefont
  {J.}~\bibnamefont {Bloch}}, \bibinfo {author} {\bibfnamefont
  {C.}~\bibnamefont {Ciuti}}, \bibinfo {author} {\bibfnamefont
  {E.}~\bibnamefont {Giacobino}},\ and\ \bibinfo {author} {\bibfnamefont
  {A.}~\bibnamefont {Bramati}},\ }\bibfield  {title} {\bibinfo {title}
  {Polariton-generated intensity squeezing in semiconductor micropillars},\
  }\href {https://doi.org/10.1038/ncomms4260} {\bibfield  {journal} {\bibinfo
  {journal} {Nature Communications}\ }\textbf {\bibinfo {volume} {5}},\
  \bibinfo {pages} {3260} (\bibinfo {year} {2014})}\BibitemShut {NoStop}%
\bibitem [{\citenamefont {Delteil}\ \emph {et~al.}(2019)\citenamefont
  {Delteil}, \citenamefont {Fink}, \citenamefont {Schade}, \citenamefont
  {H{\"o}fling}, \citenamefont {Schneider},\ and\ \citenamefont
  {{\.{I}}mamo{\u{g}}lu}}]{Delteil_NatMat2019}%
  \BibitemOpen
  \bibfield  {author} {\bibinfo {author} {\bibfnamefont {A.}~\bibnamefont
  {Delteil}}, \bibinfo {author} {\bibfnamefont {T.}~\bibnamefont {Fink}},
  \bibinfo {author} {\bibfnamefont {A.}~\bibnamefont {Schade}}, \bibinfo
  {author} {\bibfnamefont {S.}~\bibnamefont {H{\"o}fling}}, \bibinfo {author}
  {\bibfnamefont {C.}~\bibnamefont {Schneider}},\ and\ \bibinfo {author}
  {\bibfnamefont {A.}~\bibnamefont {{\.{I}}mamo{\u{g}}lu}},\ }\bibfield
  {title} {\bibinfo {title} {Towards polariton blockade of confined
  exciton--polaritons},\ }\href {https://doi.org/10.1038/s41563-019-0282-y}
  {\bibfield  {journal} {\bibinfo  {journal} {Nature Materials}\ }\textbf
  {\bibinfo {volume} {18}},\ \bibinfo {pages} {219} (\bibinfo {year}
  {2019})}\BibitemShut {NoStop}%
\bibitem [{\citenamefont {Mu{\~{n}}oz-Matutano}\ \emph
  {et~al.}(2019)\citenamefont {Mu{\~{n}}oz-Matutano}, \citenamefont {Wood},
  \citenamefont {Johnsson}, \citenamefont {Vidal}, \citenamefont {Baragiola},
  \citenamefont {Reinhard}, \citenamefont {Lema{\^i}tre}, \citenamefont
  {Bloch}, \citenamefont {Amo}, \citenamefont {Nogues}, \citenamefont {Besga},
  \citenamefont {Richard},\ and\ \citenamefont {Volz}}]{Munoz_NatMat2019}%
  \BibitemOpen
  \bibfield  {author} {\bibinfo {author} {\bibfnamefont {G.}~\bibnamefont
  {Mu{\~{n}}oz-Matutano}}, \bibinfo {author} {\bibfnamefont {A.}~\bibnamefont
  {Wood}}, \bibinfo {author} {\bibfnamefont {M.}~\bibnamefont {Johnsson}},
  \bibinfo {author} {\bibfnamefont {X.}~\bibnamefont {Vidal}}, \bibinfo
  {author} {\bibfnamefont {B.~Q.}\ \bibnamefont {Baragiola}}, \bibinfo {author}
  {\bibfnamefont {A.}~\bibnamefont {Reinhard}}, \bibinfo {author}
  {\bibfnamefont {A.}~\bibnamefont {Lema{\^i}tre}}, \bibinfo {author}
  {\bibfnamefont {J.}~\bibnamefont {Bloch}}, \bibinfo {author} {\bibfnamefont
  {A.}~\bibnamefont {Amo}}, \bibinfo {author} {\bibfnamefont {G.}~\bibnamefont
  {Nogues}}, \bibinfo {author} {\bibfnamefont {B.}~\bibnamefont {Besga}},
  \bibinfo {author} {\bibfnamefont {M.}~\bibnamefont {Richard}},\ and\ \bibinfo
  {author} {\bibfnamefont {T.}~\bibnamefont {Volz}},\ }\bibfield  {title}
  {\bibinfo {title} {Emergence of quantum correlations from interacting
  fibre-cavity polaritons},\ }\href {https://doi.org/10.1038/s41563-019-0281-z}
  {\bibfield  {journal} {\bibinfo  {journal} {Nature Materials}\ }\textbf
  {\bibinfo {volume} {18}},\ \bibinfo {pages} {213} (\bibinfo {year}
  {2019})}\BibitemShut {NoStop}%
\bibitem [{\citenamefont {Álvaro Cuevas}\ \emph {et~al.}(2018)\citenamefont
  {Álvaro Cuevas}, \citenamefont {Carreño}, \citenamefont {Silva},
  \citenamefont {Giorgi}, \citenamefont {Suárez-Forero}, \citenamefont
  {Muñoz}, \citenamefont {Fieramosca}, \citenamefont {Cardano}, \citenamefont
  {Marrucci}, \citenamefont {Tasco}, \citenamefont {Biasiol}, \citenamefont
  {del Valle}, \citenamefont {Dominici}, \citenamefont {Ballarini},
  \citenamefont {Gigli}, \citenamefont {Mataloni}, \citenamefont {Laussy},
  \citenamefont {Sciarrino},\ and\ \citenamefont
  {Sanvitto}}]{Cuevas_SciAdv2018}%
  \BibitemOpen
  \bibfield  {author} {\bibinfo {author} {\bibnamefont {Álvaro Cuevas}},
  \bibinfo {author} {\bibfnamefont {J.~C.~L.}\ \bibnamefont {Carreño}},
  \bibinfo {author} {\bibfnamefont {B.}~\bibnamefont {Silva}}, \bibinfo
  {author} {\bibfnamefont {M.~D.}\ \bibnamefont {Giorgi}}, \bibinfo {author}
  {\bibfnamefont {D.~G.}\ \bibnamefont {Suárez-Forero}}, \bibinfo {author}
  {\bibfnamefont {C.~S.}\ \bibnamefont {Muñoz}}, \bibinfo {author}
  {\bibfnamefont {A.}~\bibnamefont {Fieramosca}}, \bibinfo {author}
  {\bibfnamefont {F.}~\bibnamefont {Cardano}}, \bibinfo {author} {\bibfnamefont
  {L.}~\bibnamefont {Marrucci}}, \bibinfo {author} {\bibfnamefont
  {V.}~\bibnamefont {Tasco}}, \bibinfo {author} {\bibfnamefont
  {G.}~\bibnamefont {Biasiol}}, \bibinfo {author} {\bibfnamefont
  {E.}~\bibnamefont {del Valle}}, \bibinfo {author} {\bibfnamefont
  {L.}~\bibnamefont {Dominici}}, \bibinfo {author} {\bibfnamefont
  {D.}~\bibnamefont {Ballarini}}, \bibinfo {author} {\bibfnamefont
  {G.}~\bibnamefont {Gigli}}, \bibinfo {author} {\bibfnamefont
  {P.}~\bibnamefont {Mataloni}}, \bibinfo {author} {\bibfnamefont {F.~P.}\
  \bibnamefont {Laussy}}, \bibinfo {author} {\bibfnamefont {F.}~\bibnamefont
  {Sciarrino}},\ and\ \bibinfo {author} {\bibfnamefont {D.}~\bibnamefont
  {Sanvitto}},\ }\bibfield  {title} {\bibinfo {title} {First observation of the
  quantized exciton-polariton field and effect of interactions on a single
  polariton},\ }\href {https://doi.org/10.1126/sciadv.aao6814} {\bibfield
  {journal} {\bibinfo  {journal} {Science Advances}\ }\textbf {\bibinfo
  {volume} {4}},\ \bibinfo {pages} {eaao6814} (\bibinfo {year}
  {2018})}\BibitemShut {NoStop}%
\bibitem [{\citenamefont {Kyriienko}\ and\ \citenamefont
  {Liew}(2016)}]{Kyriienko2016}%
  \BibitemOpen
  \bibfield  {author} {\bibinfo {author} {\bibfnamefont {O.}~\bibnamefont
  {Kyriienko}}\ and\ \bibinfo {author} {\bibfnamefont {T.~C.~H.}\ \bibnamefont
  {Liew}},\ }\bibfield  {title} {\bibinfo {title} {Exciton-polariton quantum
  gates based on continuous variables},\ }\href
  {https://doi.org/10.1103/PhysRevB.93.035301} {\bibfield  {journal} {\bibinfo
  {journal} {Phys. Rev. B}\ }\textbf {\bibinfo {volume} {93}},\ \bibinfo
  {pages} {035301} (\bibinfo {year} {2016})}\BibitemShut {NoStop}%
\bibitem [{\citenamefont {Puri}\ \emph {et~al.}(2017)\citenamefont {Puri},
  \citenamefont {McMahon},\ and\ \citenamefont {Yamamoto}}]{Puri_PRB2017}%
  \BibitemOpen
  \bibfield  {author} {\bibinfo {author} {\bibfnamefont {S.}~\bibnamefont
  {Puri}}, \bibinfo {author} {\bibfnamefont {P.~L.}\ \bibnamefont {McMahon}},\
  and\ \bibinfo {author} {\bibfnamefont {Y.}~\bibnamefont {Yamamoto}},\
  }\bibfield  {title} {\bibinfo {title} {Universal logic gates for quantum-dot
  electron-spin qubits using trapped quantum-well exciton polaritons},\ }\href
  {https://doi.org/10.1103/PhysRevB.95.125410} {\bibfield  {journal} {\bibinfo
  {journal} {Phys. Rev. B}\ }\textbf {\bibinfo {volume} {95}},\ \bibinfo
  {pages} {125410} (\bibinfo {year} {2017})}\BibitemShut {NoStop}%
\bibitem [{\citenamefont {Xu}\ \emph {et~al.}(2021)\citenamefont {Xu},
  \citenamefont {Krisnanda}, \citenamefont {Verstraelen}, \citenamefont
  {Liew},\ and\ \citenamefont {Ghosh}}]{Xu_PRB2021}%
  \BibitemOpen
  \bibfield  {author} {\bibinfo {author} {\bibfnamefont {H.}~\bibnamefont
  {Xu}}, \bibinfo {author} {\bibfnamefont {T.}~\bibnamefont {Krisnanda}},
  \bibinfo {author} {\bibfnamefont {W.}~\bibnamefont {Verstraelen}}, \bibinfo
  {author} {\bibfnamefont {T.~C.~H.}\ \bibnamefont {Liew}},\ and\ \bibinfo
  {author} {\bibfnamefont {S.}~\bibnamefont {Ghosh}},\ }\bibfield  {title}
  {\bibinfo {title} {Superpolynomial quantum enhancement in polaritonic
  neuromorphic computing},\ }\href
  {https://doi.org/10.1103/PhysRevB.103.195302} {\bibfield  {journal} {\bibinfo
   {journal} {Phys. Rev. B}\ }\textbf {\bibinfo {volume} {103}},\ \bibinfo
  {pages} {195302} (\bibinfo {year} {2021})}\BibitemShut {NoStop}%
\bibitem [{\citenamefont {Nigro}\ \emph {et~al.}(2022)\citenamefont {Nigro},
  \citenamefont {D'Ambrosio}, \citenamefont {Sanvitto},\ and\ \citenamefont
  {Gerace}}]{Nigro_CommPhys2022}%
  \BibitemOpen
  \bibfield  {author} {\bibinfo {author} {\bibfnamefont {D.}~\bibnamefont
  {Nigro}}, \bibinfo {author} {\bibfnamefont {V.}~\bibnamefont {D'Ambrosio}},
  \bibinfo {author} {\bibfnamefont {D.}~\bibnamefont {Sanvitto}},\ and\
  \bibinfo {author} {\bibfnamefont {D.}~\bibnamefont {Gerace}},\ }\bibfield
  {title} {\bibinfo {title} {Integrated quantum polariton interferometry},\
  }\href {https://doi.org/10.1038/s42005-022-00810-9} {\bibfield  {journal}
  {\bibinfo  {journal} {Communications Physics}\ }\textbf {\bibinfo {volume}
  {5}},\ \bibinfo {pages} {34} (\bibinfo {year} {2022})}\BibitemShut {NoStop}%
\bibitem [{\citenamefont {Demirchyan}\ \emph {et~al.}(2014)\citenamefont
  {Demirchyan}, \citenamefont {Chestnov}, \citenamefont {Alodjants},
  \citenamefont {Glazov},\ and\ \citenamefont {Kavokin}}]{Demirchyan_PRL2014}%
  \BibitemOpen
  \bibfield  {author} {\bibinfo {author} {\bibfnamefont {S.~S.}\ \bibnamefont
  {Demirchyan}}, \bibinfo {author} {\bibfnamefont {I.~Y.}\ \bibnamefont
  {Chestnov}}, \bibinfo {author} {\bibfnamefont {A.~P.}\ \bibnamefont
  {Alodjants}}, \bibinfo {author} {\bibfnamefont {M.~M.}\ \bibnamefont
  {Glazov}},\ and\ \bibinfo {author} {\bibfnamefont {A.~V.}\ \bibnamefont
  {Kavokin}},\ }\bibfield  {title} {\bibinfo {title} {Qubits based on polariton
  rabi oscillators},\ }\href {https://doi.org/10.1103/PhysRevLett.112.196403}
  {\bibfield  {journal} {\bibinfo  {journal} {Phys. Rev. Lett.}\ }\textbf
  {\bibinfo {volume} {112}},\ \bibinfo {pages} {196403} (\bibinfo {year}
  {2014})}\BibitemShut {NoStop}%
\bibitem [{\citenamefont {Solnyshkov}\ \emph {et~al.}(2015)\citenamefont
  {Solnyshkov}, \citenamefont {Bleu},\ and\ \citenamefont
  {Malpuech}}]{Solnyshkov_SupLatt2015}%
  \BibitemOpen
  \bibfield  {author} {\bibinfo {author} {\bibfnamefont {D.}~\bibnamefont
  {Solnyshkov}}, \bibinfo {author} {\bibfnamefont {O.}~\bibnamefont {Bleu}},\
  and\ \bibinfo {author} {\bibfnamefont {G.}~\bibnamefont {Malpuech}},\
  }\bibfield  {title} {\bibinfo {title} {All optical controlled-not gate based
  on an exciton–polariton circuit},\ }\href
  {https://doi.org/https://doi.org/10.1016/j.spmi.2015.03.057} {\bibfield
  {journal} {\bibinfo  {journal} {Superlattices and Microstructures}\ }\textbf
  {\bibinfo {volume} {83}},\ \bibinfo {pages} {466} (\bibinfo {year}
  {2015})}\BibitemShut {NoStop}%
\bibitem [{\citenamefont {Ghosh}\ and\ \citenamefont {Liew}(2020)}]{Ghosh2020}%
  \BibitemOpen
  \bibfield  {author} {\bibinfo {author} {\bibfnamefont {S.}~\bibnamefont
  {Ghosh}}\ and\ \bibinfo {author} {\bibfnamefont {T.~C.~H.}\ \bibnamefont
  {Liew}},\ }\bibfield  {title} {\bibinfo {title} {Quantum computing with
  exciton-polariton condensates},\ }\href
  {https://doi.org/10.1038/s41534-020-0244-x} {\bibfield  {journal} {\bibinfo
  {journal} {Npj Quantum Inf.}\ }\textbf {\bibinfo {volume} {6}},\ \bibinfo
  {pages} {1} (\bibinfo {year} {2020})}\BibitemShut {NoStop}%
\bibitem [{\citenamefont {Xue}\ \emph {et~al.}(2021)\citenamefont {Xue},
  \citenamefont {Chestnov}, \citenamefont {Sedov}, \citenamefont {Kiktenko},
  \citenamefont {Fedorov}, \citenamefont {Schumacher}, \citenamefont {Ma},\
  and\ \citenamefont {Kavokin}}]{Xue2021}%
  \BibitemOpen
  \bibfield  {author} {\bibinfo {author} {\bibfnamefont {Y.}~\bibnamefont
  {Xue}}, \bibinfo {author} {\bibfnamefont {I.}~\bibnamefont {Chestnov}},
  \bibinfo {author} {\bibfnamefont {E.}~\bibnamefont {Sedov}}, \bibinfo
  {author} {\bibfnamefont {E.}~\bibnamefont {Kiktenko}}, \bibinfo {author}
  {\bibfnamefont {A.~K.}\ \bibnamefont {Fedorov}}, \bibinfo {author}
  {\bibfnamefont {S.}~\bibnamefont {Schumacher}}, \bibinfo {author}
  {\bibfnamefont {X.}~\bibnamefont {Ma}},\ and\ \bibinfo {author}
  {\bibfnamefont {A.}~\bibnamefont {Kavokin}},\ }\bibfield  {title} {\bibinfo
  {title} {Split-ring polariton condensates as macroscopic two-level quantum
  systems},\ }\href {https://doi.org/10.1103/PhysRevResearch.3.013099}
  {\bibfield  {journal} {\bibinfo  {journal} {Phys. Rev. Research}\ }\textbf
  {\bibinfo {volume} {3}},\ \bibinfo {pages} {013099} (\bibinfo {year}
  {2021})}\BibitemShut {NoStop}%
\bibitem [{\citenamefont {Carusotto}\ and\ \citenamefont
  {Ciuti}(2013)}]{Carusotto2013}%
  \BibitemOpen
  \bibfield  {author} {\bibinfo {author} {\bibfnamefont {I.}~\bibnamefont
  {Carusotto}}\ and\ \bibinfo {author} {\bibfnamefont {C.}~\bibnamefont
  {Ciuti}},\ }\bibfield  {title} {\bibinfo {title} {Quantum fluids of light},\
  }\href {https://doi.org/10.1103/RevModPhys.85.299} {\bibfield  {journal}
  {\bibinfo  {journal} {Rev. Mod. Phys.}\ }\textbf {\bibinfo {volume} {85}},\
  \bibinfo {pages} {299} (\bibinfo {year} {2013})}\BibitemShut {NoStop}%
\bibitem [{\citenamefont {Kasprzak}\ \emph {et~al.}(2006)\citenamefont
  {Kasprzak}, \citenamefont {Richard}, \citenamefont {Kundermann},
  \citenamefont {Baas}, \citenamefont {Jeambrun}, \citenamefont {Keeling},
  \citenamefont {Marchetti}, \citenamefont {Szyma{\'n}ska}, \citenamefont
  {Andr{\'e}}, \citenamefont {Staehli}, \citenamefont {Savona}, \citenamefont
  {Littlewood}, \citenamefont {Deveaud},\ and\ \citenamefont
  {Dang}}]{Kasprzak2006}%
  \BibitemOpen
  \bibfield  {author} {\bibinfo {author} {\bibfnamefont {J.}~\bibnamefont
  {Kasprzak}}, \bibinfo {author} {\bibfnamefont {M.}~\bibnamefont {Richard}},
  \bibinfo {author} {\bibfnamefont {S.}~\bibnamefont {Kundermann}}, \bibinfo
  {author} {\bibfnamefont {A.}~\bibnamefont {Baas}}, \bibinfo {author}
  {\bibfnamefont {P.}~\bibnamefont {Jeambrun}}, \bibinfo {author}
  {\bibfnamefont {J.}~\bibnamefont {Keeling}}, \bibinfo {author} {\bibfnamefont
  {F.}~\bibnamefont {Marchetti}}, \bibinfo {author} {\bibfnamefont
  {M.}~\bibnamefont {Szyma{\'n}ska}}, \bibinfo {author} {\bibfnamefont
  {R.}~\bibnamefont {Andr{\'e}}}, \bibinfo {author} {\bibfnamefont
  {J.}~\bibnamefont {Staehli}}, \bibinfo {author} {\bibfnamefont
  {V.}~\bibnamefont {Savona}}, \bibinfo {author} {\bibfnamefont
  {P.}~\bibnamefont {Littlewood}}, \bibinfo {author} {\bibfnamefont
  {B.}~\bibnamefont {Deveaud}},\ and\ \bibinfo {author} {\bibfnamefont {L.~S.}\
  \bibnamefont {Dang}},\ }\bibfield  {title} {\bibinfo {title}
  {{B}ose--{E}instein condensation of exciton polaritons},\ }\href@noop {}
  {\bibfield  {journal} {\bibinfo  {journal} {Nature}\ }\textbf {\bibinfo
  {volume} {443}},\ \bibinfo {pages} {409} (\bibinfo {year}
  {2006})}\BibitemShut {NoStop}%
\bibitem [{\citenamefont {Lukoshkin}\ \emph {et~al.}(2018)\citenamefont
  {Lukoshkin}, \citenamefont {Kalevich}, \citenamefont {Afanasiev},
  \citenamefont {Kavokin}, \citenamefont {Hatzopoulos}, \citenamefont
  {Savvidis}, \citenamefont {Sedov},\ and\ \citenamefont
  {Kavokin}}]{Lukoshkin_PRB2018}%
  \BibitemOpen
  \bibfield  {author} {\bibinfo {author} {\bibfnamefont {V.~A.}\ \bibnamefont
  {Lukoshkin}}, \bibinfo {author} {\bibfnamefont {V.~K.}\ \bibnamefont
  {Kalevich}}, \bibinfo {author} {\bibfnamefont {M.~M.}\ \bibnamefont
  {Afanasiev}}, \bibinfo {author} {\bibfnamefont {K.~V.}\ \bibnamefont
  {Kavokin}}, \bibinfo {author} {\bibfnamefont {Z.}~\bibnamefont
  {Hatzopoulos}}, \bibinfo {author} {\bibfnamefont {P.~G.}\ \bibnamefont
  {Savvidis}}, \bibinfo {author} {\bibfnamefont {E.~S.}\ \bibnamefont
  {Sedov}},\ and\ \bibinfo {author} {\bibfnamefont {A.~V.}\ \bibnamefont
  {Kavokin}},\ }\bibfield  {title} {\bibinfo {title} {Persistent circular
  currents of exciton-polaritons in cylindrical pillar microcavities},\ }\href
  {https://doi.org/10.1103/PhysRevB.97.195149} {\bibfield  {journal} {\bibinfo
  {journal} {Phys. Rev. B}\ }\textbf {\bibinfo {volume} {97}},\ \bibinfo
  {pages} {195149} (\bibinfo {year} {2018})}\BibitemShut {NoStop}%
\bibitem [{\citenamefont {Sedov}\ \emph {et~al.}(2020)\citenamefont {Sedov},
  \citenamefont {Lukoshkin}, \citenamefont {Kalevich}, \citenamefont
  {Hatzopoulos}, \citenamefont {Savvidis},\ and\ \citenamefont
  {Kavokin}}]{Sedov_ACSPhot2020}%
  \BibitemOpen
  \bibfield  {author} {\bibinfo {author} {\bibfnamefont {E.}~\bibnamefont
  {Sedov}}, \bibinfo {author} {\bibfnamefont {V.}~\bibnamefont {Lukoshkin}},
  \bibinfo {author} {\bibfnamefont {V.}~\bibnamefont {Kalevich}}, \bibinfo
  {author} {\bibfnamefont {Z.}~\bibnamefont {Hatzopoulos}}, \bibinfo {author}
  {\bibfnamefont {P.}~\bibnamefont {Savvidis}},\ and\ \bibinfo {author}
  {\bibfnamefont {A.}~\bibnamefont {Kavokin}},\ }\bibfield  {title} {\bibinfo
  {title} {Persistent currents in half-moon polariton condensates},\ }\href
  {https://doi.org/10.1021/acsphotonics.9b01779} {\bibfield  {journal}
  {\bibinfo  {journal} {ACS Photonics}\ }\textbf {\bibinfo {volume} {7}},\
  \bibinfo {pages} {1163} (\bibinfo {year} {2020})}\BibitemShut {NoStop}%
\bibitem [{\citenamefont {Sedov}\ \emph {et~al.}(2021)\citenamefont {Sedov},
  \citenamefont {Lukoshkin}, \citenamefont {Kalevich}, \citenamefont
  {Savvidis},\ and\ \citenamefont {Kavokin}}]{Sedov2021}%
  \BibitemOpen
  \bibfield  {author} {\bibinfo {author} {\bibfnamefont {E.~S.}\ \bibnamefont
  {Sedov}}, \bibinfo {author} {\bibfnamefont {V.~A.}\ \bibnamefont
  {Lukoshkin}}, \bibinfo {author} {\bibfnamefont {V.~K.}\ \bibnamefont
  {Kalevich}}, \bibinfo {author} {\bibfnamefont {P.~G.}\ \bibnamefont
  {Savvidis}},\ and\ \bibinfo {author} {\bibfnamefont {A.~V.}\ \bibnamefont
  {Kavokin}},\ }\bibfield  {title} {\bibinfo {title} {Circular polariton
  currents with integer and fractional orbital angular momenta},\ }\href
  {https://doi.org/10.1103/PhysRevResearch.3.013072} {\bibfield  {journal}
  {\bibinfo  {journal} {Phys. Rev. Research}\ }\textbf {\bibinfo {volume}
  {3}},\ \bibinfo {pages} {013072} (\bibinfo {year} {2021})}\BibitemShut
  {NoStop}%
\bibitem [{\citenamefont {Real}\ \emph {et~al.}(2021)\citenamefont {Real},
  \citenamefont {Carlon~Zambon}, \citenamefont {St-Jean}, \citenamefont
  {Sagnes}, \citenamefont {Lema\^{\i}tre}, \citenamefont {Le~Gratiet},
  \citenamefont {Harouri}, \citenamefont {Ravets}, \citenamefont {Bloch},\ and\
  \citenamefont {Amo}}]{Real_PRR2021}%
  \BibitemOpen
  \bibfield  {author} {\bibinfo {author} {\bibfnamefont {B.}~\bibnamefont
  {Real}}, \bibinfo {author} {\bibfnamefont {N.}~\bibnamefont {Carlon~Zambon}},
  \bibinfo {author} {\bibfnamefont {P.}~\bibnamefont {St-Jean}}, \bibinfo
  {author} {\bibfnamefont {I.}~\bibnamefont {Sagnes}}, \bibinfo {author}
  {\bibfnamefont {A.}~\bibnamefont {Lema\^{\i}tre}}, \bibinfo {author}
  {\bibfnamefont {L.}~\bibnamefont {Le~Gratiet}}, \bibinfo {author}
  {\bibfnamefont {A.}~\bibnamefont {Harouri}}, \bibinfo {author} {\bibfnamefont
  {S.}~\bibnamefont {Ravets}}, \bibinfo {author} {\bibfnamefont
  {J.}~\bibnamefont {Bloch}},\ and\ \bibinfo {author} {\bibfnamefont
  {A.}~\bibnamefont {Amo}},\ }\bibfield  {title} {\bibinfo {title} {Chiral
  emission induced by optical zeeman effect in polariton micropillars},\ }\href
  {https://doi.org/10.1103/PhysRevResearch.3.043161} {\bibfield  {journal}
  {\bibinfo  {journal} {Phys. Rev. Res.}\ }\textbf {\bibinfo {volume} {3}},\
  \bibinfo {pages} {043161} (\bibinfo {year} {2021})}\BibitemShut {NoStop}%
\bibitem [{\citenamefont {Lukoshkin}\ \emph {et~al.}(2023)\citenamefont
  {Lukoshkin}, \citenamefont {Sedov}, \citenamefont {Kalevich}, \citenamefont
  {Hatzopoulos}, \citenamefont {Savvidis},\ and\ \citenamefont
  {Kavokin}}]{Lukoshkin_SciRep2023}%
  \BibitemOpen
  \bibfield  {author} {\bibinfo {author} {\bibfnamefont {V.}~\bibnamefont
  {Lukoshkin}}, \bibinfo {author} {\bibfnamefont {E.}~\bibnamefont {Sedov}},
  \bibinfo {author} {\bibfnamefont {V.}~\bibnamefont {Kalevich}}, \bibinfo
  {author} {\bibfnamefont {Z.}~\bibnamefont {Hatzopoulos}}, \bibinfo {author}
  {\bibfnamefont {P.~G.}\ \bibnamefont {Savvidis}},\ and\ \bibinfo {author}
  {\bibfnamefont {A.}~\bibnamefont {Kavokin}},\ }\bibfield  {title} {\bibinfo
  {title} {Steady state oscillations of circular currents in concentric
  polariton condensates},\ }\href {https://doi.org/10.1038/s41598-023-31520-z}
  {\bibfield  {journal} {\bibinfo  {journal} {Scientific Reports}\ }\textbf
  {\bibinfo {volume} {13}},\ \bibinfo {pages} {4607} (\bibinfo {year}
  {2023})}\BibitemShut {NoStop}%
\bibitem [{\citenamefont {Nardin}\ \emph {et~al.}(2010)\citenamefont {Nardin},
  \citenamefont {Lagoudakis}, \citenamefont {Pietka}, \citenamefont
  {Morier-Genoud}, \citenamefont {L\'eger},\ and\ \citenamefont
  {Deveaud-Pl\'edran}}]{NardinPhysRevB82073303(2010)}%
  \BibitemOpen
  \bibfield  {author} {\bibinfo {author} {\bibfnamefont {G.}~\bibnamefont
  {Nardin}}, \bibinfo {author} {\bibfnamefont {K.~G.}\ \bibnamefont
  {Lagoudakis}}, \bibinfo {author} {\bibfnamefont {B.}~\bibnamefont {Pietka}},
  \bibinfo {author} {\bibfnamefont {F.~m.~c.}\ \bibnamefont {Morier-Genoud}},
  \bibinfo {author} {\bibfnamefont {Y.}~\bibnamefont {L\'eger}},\ and\ \bibinfo
  {author} {\bibfnamefont {B.}~\bibnamefont {Deveaud-Pl\'edran}},\ }\bibfield
  {title} {\bibinfo {title} {Selective photoexcitation of confined
  exciton-polariton vortices},\ }\href
  {https://doi.org/10.1103/PhysRevB.82.073303} {\bibfield  {journal} {\bibinfo
  {journal} {Phys. Rev. B}\ }\textbf {\bibinfo {volume} {82}},\ \bibinfo
  {pages} {073303} (\bibinfo {year} {2010})}\BibitemShut {NoStop}%
\bibitem [{\citenamefont {Gao}\ \emph {et~al.}(2018)\citenamefont {Gao},
  \citenamefont {Egorov}, \citenamefont {Estrecho}, \citenamefont {Winkler},
  \citenamefont {Kamp}, \citenamefont {Schneider}, \citenamefont {H\"ofling},
  \citenamefont {Truscott},\ and\ \citenamefont {Ostrovskaya}}]{Gao_PRL2018}%
  \BibitemOpen
  \bibfield  {author} {\bibinfo {author} {\bibfnamefont {T.}~\bibnamefont
  {Gao}}, \bibinfo {author} {\bibfnamefont {O.~A.}\ \bibnamefont {Egorov}},
  \bibinfo {author} {\bibfnamefont {E.}~\bibnamefont {Estrecho}}, \bibinfo
  {author} {\bibfnamefont {K.}~\bibnamefont {Winkler}}, \bibinfo {author}
  {\bibfnamefont {M.}~\bibnamefont {Kamp}}, \bibinfo {author} {\bibfnamefont
  {C.}~\bibnamefont {Schneider}}, \bibinfo {author} {\bibfnamefont
  {S.}~\bibnamefont {H\"ofling}}, \bibinfo {author} {\bibfnamefont {A.~G.}\
  \bibnamefont {Truscott}},\ and\ \bibinfo {author} {\bibfnamefont {E.~A.}\
  \bibnamefont {Ostrovskaya}},\ }\bibfield  {title} {\bibinfo {title}
  {Controlled ordering of topological charges in an exciton-polariton chain},\
  }\href {https://doi.org/10.1103/PhysRevLett.121.225302} {\bibfield  {journal}
  {\bibinfo  {journal} {Phys. Rev. Lett.}\ }\textbf {\bibinfo {volume} {121}},\
  \bibinfo {pages} {225302} (\bibinfo {year} {2018})}\BibitemShut {NoStop}%
\bibitem [{\citenamefont {Kim}\ \emph {et~al.}(2011)\citenamefont {Kim},
  \citenamefont {Kusudo}, \citenamefont {Wu}, \citenamefont {Masumoto},
  \citenamefont {L{\"o}ffler}, \citenamefont {H{\"o}fling}, \citenamefont
  {Kumada}, \citenamefont {Worschech}, \citenamefont {Forchel},\ and\
  \citenamefont {Yamamoto}}]{Kim_NatPhys2011}%
  \BibitemOpen
  \bibfield  {author} {\bibinfo {author} {\bibfnamefont {N.~Y.}\ \bibnamefont
  {Kim}}, \bibinfo {author} {\bibfnamefont {K.}~\bibnamefont {Kusudo}},
  \bibinfo {author} {\bibfnamefont {C.}~\bibnamefont {Wu}}, \bibinfo {author}
  {\bibfnamefont {N.}~\bibnamefont {Masumoto}}, \bibinfo {author}
  {\bibfnamefont {A.}~\bibnamefont {L{\"o}ffler}}, \bibinfo {author}
  {\bibfnamefont {S.}~\bibnamefont {H{\"o}fling}}, \bibinfo {author}
  {\bibfnamefont {N.}~\bibnamefont {Kumada}}, \bibinfo {author} {\bibfnamefont
  {L.}~\bibnamefont {Worschech}}, \bibinfo {author} {\bibfnamefont
  {A.}~\bibnamefont {Forchel}},\ and\ \bibinfo {author} {\bibfnamefont
  {Y.}~\bibnamefont {Yamamoto}},\ }\bibfield  {title} {\bibinfo {title}
  {Dynamical d-wave condensation of exciton--polaritons in a two-dimensional
  square-lattice potential},\ }\href {https://doi.org/10.1038/nphys2012}
  {\bibfield  {journal} {\bibinfo  {journal} {Nature Physics}\ }\textbf
  {\bibinfo {volume} {7}},\ \bibinfo {pages} {681} (\bibinfo {year}
  {2011})}\BibitemShut {NoStop}%
\bibitem [{\citenamefont {Askitopoulos}\ \emph {et~al.}(2015)\citenamefont
  {Askitopoulos}, \citenamefont {Liew}, \citenamefont {Ohadi}, \citenamefont
  {Hatzopoulos}, \citenamefont {Savvidis},\ and\ \citenamefont
  {Lagoudakis}}]{Askitopoulos2015}%
  \BibitemOpen
  \bibfield  {author} {\bibinfo {author} {\bibfnamefont {A.}~\bibnamefont
  {Askitopoulos}}, \bibinfo {author} {\bibfnamefont {T.~C.~H.}\ \bibnamefont
  {Liew}}, \bibinfo {author} {\bibfnamefont {H.}~\bibnamefont {Ohadi}},
  \bibinfo {author} {\bibfnamefont {Z.}~\bibnamefont {Hatzopoulos}}, \bibinfo
  {author} {\bibfnamefont {P.~G.}\ \bibnamefont {Savvidis}},\ and\ \bibinfo
  {author} {\bibfnamefont {P.~G.}\ \bibnamefont {Lagoudakis}},\ }\bibfield
  {title} {\bibinfo {title} {Robust platform for engineering pure-quantum-state
  transitions in polariton condensates},\ }\href
  {https://doi.org/10.1103/PhysRevB.92.035305} {\bibfield  {journal} {\bibinfo
  {journal} {Phys. Rev. B}\ }\textbf {\bibinfo {volume} {92}},\ \bibinfo
  {pages} {035305} (\bibinfo {year} {2015})}\BibitemShut {NoStop}%
\bibitem [{\citenamefont {Dall}\ \emph {et~al.}(2014)\citenamefont {Dall},
  \citenamefont {Fraser}, \citenamefont {Desyatnikov}, \citenamefont {Li},
  \citenamefont {Brodbeck}, \citenamefont {Kamp}, \citenamefont {Schneider},
  \citenamefont {H\"ofling},\ and\ \citenamefont {Ostrovskaya}}]{Dall_PRL2014}%
  \BibitemOpen
  \bibfield  {author} {\bibinfo {author} {\bibfnamefont {R.}~\bibnamefont
  {Dall}}, \bibinfo {author} {\bibfnamefont {M.~D.}\ \bibnamefont {Fraser}},
  \bibinfo {author} {\bibfnamefont {A.~S.}\ \bibnamefont {Desyatnikov}},
  \bibinfo {author} {\bibfnamefont {G.}~\bibnamefont {Li}}, \bibinfo {author}
  {\bibfnamefont {S.}~\bibnamefont {Brodbeck}}, \bibinfo {author}
  {\bibfnamefont {M.}~\bibnamefont {Kamp}}, \bibinfo {author} {\bibfnamefont
  {C.}~\bibnamefont {Schneider}}, \bibinfo {author} {\bibfnamefont
  {S.}~\bibnamefont {H\"ofling}},\ and\ \bibinfo {author} {\bibfnamefont
  {E.~A.}\ \bibnamefont {Ostrovskaya}},\ }\bibfield  {title} {\bibinfo {title}
  {Creation of orbital angular momentum states with chiral polaritonic
  lenses},\ }\href {https://doi.org/10.1103/PhysRevLett.113.200404} {\bibfield
  {journal} {\bibinfo  {journal} {Phys. Rev. Lett.}\ }\textbf {\bibinfo
  {volume} {113}},\ \bibinfo {pages} {200404} (\bibinfo {year}
  {2014})}\BibitemShut {NoStop}%
\bibitem [{\citenamefont {Sun}\ \emph {et~al.}(2018)\citenamefont {Sun},
  \citenamefont {Yoon}, \citenamefont {Khan}, \citenamefont {Ge}, \citenamefont
  {Steger}, \citenamefont {Pfeiffer}, \citenamefont {West}, \citenamefont
  {T\"ureci}, \citenamefont {Snoke},\ and\ \citenamefont
  {Nelson}}]{Sun_PRB2018}%
  \BibitemOpen
  \bibfield  {author} {\bibinfo {author} {\bibfnamefont {Y.}~\bibnamefont
  {Sun}}, \bibinfo {author} {\bibfnamefont {Y.}~\bibnamefont {Yoon}}, \bibinfo
  {author} {\bibfnamefont {S.}~\bibnamefont {Khan}}, \bibinfo {author}
  {\bibfnamefont {L.}~\bibnamefont {Ge}}, \bibinfo {author} {\bibfnamefont
  {M.}~\bibnamefont {Steger}}, \bibinfo {author} {\bibfnamefont {L.~N.}\
  \bibnamefont {Pfeiffer}}, \bibinfo {author} {\bibfnamefont {K.}~\bibnamefont
  {West}}, \bibinfo {author} {\bibfnamefont {H.~E.}\ \bibnamefont {T\"ureci}},
  \bibinfo {author} {\bibfnamefont {D.~W.}\ \bibnamefont {Snoke}},\ and\
  \bibinfo {author} {\bibfnamefont {K.~A.}\ \bibnamefont {Nelson}},\ }\bibfield
   {title} {\bibinfo {title} {Stable switching among high-order modes in
  polariton condensates},\ }\href {https://doi.org/10.1103/PhysRevB.97.045303}
  {\bibfield  {journal} {\bibinfo  {journal} {Phys. Rev. B}\ }\textbf {\bibinfo
  {volume} {97}},\ \bibinfo {pages} {045303} (\bibinfo {year}
  {2018})}\BibitemShut {NoStop}%
\bibitem [{\citenamefont {Askitopoulos}\ \emph {et~al.}(2018)\citenamefont
  {Askitopoulos}, \citenamefont {Nalitov}, \citenamefont {Sedov}, \citenamefont
  {Pickup}, \citenamefont {Cherotchenko}, \citenamefont {Hatzopoulos},
  \citenamefont {Savvidis}, \citenamefont {Kavokin},\ and\ \citenamefont
  {Lagoudakis}}]{Askitopoulos_PRB2018}%
  \BibitemOpen
  \bibfield  {author} {\bibinfo {author} {\bibfnamefont {A.}~\bibnamefont
  {Askitopoulos}}, \bibinfo {author} {\bibfnamefont {A.~V.}\ \bibnamefont
  {Nalitov}}, \bibinfo {author} {\bibfnamefont {E.~S.}\ \bibnamefont {Sedov}},
  \bibinfo {author} {\bibfnamefont {L.}~\bibnamefont {Pickup}}, \bibinfo
  {author} {\bibfnamefont {E.~D.}\ \bibnamefont {Cherotchenko}}, \bibinfo
  {author} {\bibfnamefont {Z.}~\bibnamefont {Hatzopoulos}}, \bibinfo {author}
  {\bibfnamefont {P.~G.}\ \bibnamefont {Savvidis}}, \bibinfo {author}
  {\bibfnamefont {A.~V.}\ \bibnamefont {Kavokin}},\ and\ \bibinfo {author}
  {\bibfnamefont {P.~G.}\ \bibnamefont {Lagoudakis}},\ }\bibfield  {title}
  {\bibinfo {title} {All-optical quantum fluid spin beam splitter},\ }\href
  {https://doi.org/10.1103/PhysRevB.97.235303} {\bibfield  {journal} {\bibinfo
  {journal} {Phys. Rev. B}\ }\textbf {\bibinfo {volume} {97}},\ \bibinfo
  {pages} {235303} (\bibinfo {year} {2018})}\BibitemShut {NoStop}%
\bibitem [{\citenamefont {T\"opfer}\ \emph {et~al.}(2020)\citenamefont
  {T\"opfer}, \citenamefont {Sigurdsson}, \citenamefont {Alyatkin},\ and\
  \citenamefont {Lagoudakis}}]{Topfer_PRB2020}%
  \BibitemOpen
  \bibfield  {author} {\bibinfo {author} {\bibfnamefont {J.~D.}\ \bibnamefont
  {T\"opfer}}, \bibinfo {author} {\bibfnamefont {H.}~\bibnamefont
  {Sigurdsson}}, \bibinfo {author} {\bibfnamefont {S.}~\bibnamefont
  {Alyatkin}},\ and\ \bibinfo {author} {\bibfnamefont {P.~G.}\ \bibnamefont
  {Lagoudakis}},\ }\bibfield  {title} {\bibinfo {title} {{L}otka-{V}olterra
  population dynamics in coherent and tunable oscillators of trapped polariton
  condensates},\ }\href {https://doi.org/10.1103/PhysRevB.102.195428}
  {\bibfield  {journal} {\bibinfo  {journal} {Phys. Rev. B}\ }\textbf {\bibinfo
  {volume} {102}},\ \bibinfo {pages} {195428} (\bibinfo {year}
  {2020})}\BibitemShut {NoStop}%
\bibitem [{\citenamefont {Sitnik}\ \emph {et~al.}(2022)\citenamefont {Sitnik},
  \citenamefont {Alyatkin}, \citenamefont {T\"opfer}, \citenamefont {Gnusov},
  \citenamefont {Cookson}, \citenamefont {Sigurdsson},\ and\ \citenamefont
  {Lagoudakis}}]{Sitnik_PRL2022}%
  \BibitemOpen
  \bibfield  {author} {\bibinfo {author} {\bibfnamefont {K.~A.}\ \bibnamefont
  {Sitnik}}, \bibinfo {author} {\bibfnamefont {S.}~\bibnamefont {Alyatkin}},
  \bibinfo {author} {\bibfnamefont {J.~D.}\ \bibnamefont {T\"opfer}}, \bibinfo
  {author} {\bibfnamefont {I.}~\bibnamefont {Gnusov}}, \bibinfo {author}
  {\bibfnamefont {T.}~\bibnamefont {Cookson}}, \bibinfo {author} {\bibfnamefont
  {H.}~\bibnamefont {Sigurdsson}},\ and\ \bibinfo {author} {\bibfnamefont
  {P.~G.}\ \bibnamefont {Lagoudakis}},\ }\bibfield  {title} {\bibinfo {title}
  {Spontaneous formation of time-periodic vortex cluster in nonlinear fluids of
  light},\ }\href {https://doi.org/10.1103/PhysRevLett.128.237402} {\bibfield
  {journal} {\bibinfo  {journal} {Phys. Rev. Lett.}\ }\textbf {\bibinfo
  {volume} {128}},\ \bibinfo {pages} {237402} (\bibinfo {year}
  {2022})}\BibitemShut {NoStop}%
\bibitem [{\citenamefont {Barrat}\ \emph {et~al.}(2023)\citenamefont {Barrat},
  \citenamefont {Tzortzakakis}, \citenamefont {Niu}, \citenamefont {Zhou},
  \citenamefont {Paschos}, \citenamefont {Petrosyan},\ and\ \citenamefont
  {Savvidis}}]{Barrat2023superfluid}%
  \BibitemOpen
  \bibfield  {author} {\bibinfo {author} {\bibfnamefont {J.}~\bibnamefont
  {Barrat}}, \bibinfo {author} {\bibfnamefont {A.~F.}\ \bibnamefont
  {Tzortzakakis}}, \bibinfo {author} {\bibfnamefont {M.}~\bibnamefont {Niu}},
  \bibinfo {author} {\bibfnamefont {X.}~\bibnamefont {Zhou}}, \bibinfo {author}
  {\bibfnamefont {G.~G.}\ \bibnamefont {Paschos}}, \bibinfo {author}
  {\bibfnamefont {D.}~\bibnamefont {Petrosyan}},\ and\ \bibinfo {author}
  {\bibfnamefont {P.~G.}\ \bibnamefont {Savvidis}},\ }\href@noop {} {\bibinfo
  {title} {Superfluid polaritonic qubit in an annular trap}} (\bibinfo {year}
  {2023}),\ \Eprint {https://arxiv.org/abs/2308.05555} {arXiv:2308.05555
  [cond-mat.quant-gas]} \BibitemShut {NoStop}%
\bibitem [{\citenamefont {Berloff}\ \emph {et~al.}(2017)\citenamefont
  {Berloff}, \citenamefont {Silva}, \citenamefont {Kalinin}, \citenamefont
  {Askitopoulos}, \citenamefont {T{\"o}pfer}, \citenamefont {Cilibrizzi},
  \citenamefont {Langbein},\ and\ \citenamefont {Lagoudakis}}]{Berloff2017}%
  \BibitemOpen
  \bibfield  {author} {\bibinfo {author} {\bibfnamefont {N.~G.}\ \bibnamefont
  {Berloff}}, \bibinfo {author} {\bibfnamefont {M.}~\bibnamefont {Silva}},
  \bibinfo {author} {\bibfnamefont {K.}~\bibnamefont {Kalinin}}, \bibinfo
  {author} {\bibfnamefont {A.}~\bibnamefont {Askitopoulos}}, \bibinfo {author}
  {\bibfnamefont {J.~D.}\ \bibnamefont {T{\"o}pfer}}, \bibinfo {author}
  {\bibfnamefont {P.}~\bibnamefont {Cilibrizzi}}, \bibinfo {author}
  {\bibfnamefont {W.}~\bibnamefont {Langbein}},\ and\ \bibinfo {author}
  {\bibfnamefont {P.~G.}\ \bibnamefont {Lagoudakis}},\ }\bibfield  {title}
  {\bibinfo {title} {Realizing the classical {XY} hamiltonian in polariton
  simulators},\ }\href {https://doi.org/10.1038/nmat4971} {\bibfield  {journal}
  {\bibinfo  {journal} {Nature Materials}\ }\textbf {\bibinfo {volume} {16}},\
  \bibinfo {pages} {1120} (\bibinfo {year} {2017})}\BibitemShut {NoStop}%
\bibitem [{\citenamefont {Harrison}\ \emph {et~al.}(2022)\citenamefont
  {Harrison}, \citenamefont {Sigurdsson}, \citenamefont {Alyatkin},
  \citenamefont {T\"opfer},\ and\ \citenamefont
  {Lagoudakis}}]{Harrison_PRAppl2022}%
  \BibitemOpen
  \bibfield  {author} {\bibinfo {author} {\bibfnamefont {S.}~\bibnamefont
  {Harrison}}, \bibinfo {author} {\bibfnamefont {H.}~\bibnamefont
  {Sigurdsson}}, \bibinfo {author} {\bibfnamefont {S.}~\bibnamefont
  {Alyatkin}}, \bibinfo {author} {\bibfnamefont {J.}~\bibnamefont {T\"opfer}},\
  and\ \bibinfo {author} {\bibfnamefont {P.}~\bibnamefont {Lagoudakis}},\
  }\bibfield  {title} {\bibinfo {title} {Solving the max-3-cut problem with
  coherent networks},\ }\href
  {https://doi.org/10.1103/PhysRevApplied.17.024063} {\bibfield  {journal}
  {\bibinfo  {journal} {Phys. Rev. Appl.}\ }\textbf {\bibinfo {volume} {17}},\
  \bibinfo {pages} {024063} (\bibinfo {year} {2022})}\BibitemShut {NoStop}%
\bibitem [{\citenamefont {Kalinin}\ \emph {et~al.}(2020)\citenamefont
  {Kalinin}, \citenamefont {Amo}, \citenamefont {Bloch},\ and\ \citenamefont
  {Berloff}}]{Kalinin_Nanopho2020}%
  \BibitemOpen
  \bibfield  {author} {\bibinfo {author} {\bibfnamefont {K.~P.}\ \bibnamefont
  {Kalinin}}, \bibinfo {author} {\bibfnamefont {A.}~\bibnamefont {Amo}},
  \bibinfo {author} {\bibfnamefont {J.}~\bibnamefont {Bloch}},\ and\ \bibinfo
  {author} {\bibfnamefont {N.~G.}\ \bibnamefont {Berloff}},\ }\bibfield
  {title} {\bibinfo {title} {Polaritonic xy-ising machine},\ }\href
  {https://doi.org/doi:10.1515/nanoph-2020-0162} {\bibfield  {journal}
  {\bibinfo  {journal} {Nanophotonics}\ }\textbf {\bibinfo {volume} {9}},\
  \bibinfo {pages} {4127} (\bibinfo {year} {2020})}\BibitemShut {NoStop}%
\bibitem [{\citenamefont {Alyatkin}\ \emph {et~al.}(2022)\citenamefont
  {Alyatkin}, \citenamefont {Milian}, \citenamefont {Kartashov}, \citenamefont
  {Sitnik}, \citenamefont {Topfer}, \citenamefont {Sigurdsson},\ and\
  \citenamefont {Lagoudakis}}]{alyatkin2022alloptical}%
  \BibitemOpen
  \bibfield  {author} {\bibinfo {author} {\bibfnamefont {S.}~\bibnamefont
  {Alyatkin}}, \bibinfo {author} {\bibfnamefont {C.}~\bibnamefont {Milian}},
  \bibinfo {author} {\bibfnamefont {Y.~V.}\ \bibnamefont {Kartashov}}, \bibinfo
  {author} {\bibfnamefont {K.~A.}\ \bibnamefont {Sitnik}}, \bibinfo {author}
  {\bibfnamefont {J.~D.}\ \bibnamefont {Topfer}}, \bibinfo {author}
  {\bibfnamefont {H.}~\bibnamefont {Sigurdsson}},\ and\ \bibinfo {author}
  {\bibfnamefont {P.~G.}\ \bibnamefont {Lagoudakis}},\ }\href@noop {} {\bibinfo
  {title} {All-optical artificial vortex matter in quantum fluids of light}}
  (\bibinfo {year} {2022}),\ \Eprint {https://arxiv.org/abs/2207.01850}
  {arXiv:2207.01850 [cond-mat.mes-hall]} \BibitemShut {NoStop}%
\bibitem [{\citenamefont {de~Oliveira}\ and\ \citenamefont
  {Munro}(2000)}]{Oliveira_PRA2000}%
  \BibitemOpen
  \bibfield  {author} {\bibinfo {author} {\bibfnamefont {M.~C.}\ \bibnamefont
  {de~Oliveira}}\ and\ \bibinfo {author} {\bibfnamefont {W.~J.}\ \bibnamefont
  {Munro}},\ }\bibfield  {title} {\bibinfo {title} {Quantum computation with
  mesoscopic superposition states},\ }\href
  {https://doi.org/10.1103/PhysRevA.61.042309} {\bibfield  {journal} {\bibinfo
  {journal} {Phys. Rev. A}\ }\textbf {\bibinfo {volume} {61}},\ \bibinfo
  {pages} {042309} (\bibinfo {year} {2000})}\BibitemShut {NoStop}%
\bibitem [{\citenamefont {Nardin}\ \emph {et~al.}(2011)\citenamefont {Nardin},
  \citenamefont {L{\'e}ger}, \citenamefont {Pietka}, \citenamefont
  {Morier-Genoud},\ and\ \citenamefont
  {Deveaud-Pledran}}]{GaelNardinJournalofNanophot2011}%
  \BibitemOpen
  \bibfield  {author} {\bibinfo {author} {\bibfnamefont {G.}~\bibnamefont
  {Nardin}}, \bibinfo {author} {\bibfnamefont {Y.}~\bibnamefont {L{\'e}ger}},
  \bibinfo {author} {\bibfnamefont {B.}~\bibnamefont {Pietka}}, \bibinfo
  {author} {\bibfnamefont {F.}~\bibnamefont {Morier-Genoud}},\ and\ \bibinfo
  {author} {\bibfnamefont {B.}~\bibnamefont {Deveaud-Pledran}},\ }\bibfield
  {title} {\bibinfo {title} {{Coherent oscillations between orbital angular
  momentum polariton states in an elliptic resonator}},\ }\href
  {https://doi.org/10.1117/1.3609825} {\bibfield  {journal} {\bibinfo
  {journal} {Journal of Nanophotonics}\ }\textbf {\bibinfo {volume} {5}},\
  \bibinfo {pages} {053517} (\bibinfo {year} {2011})}\BibitemShut {NoStop}%
\bibitem [{\citenamefont {Bennenhei}\ \emph {et~al.}(2023)\citenamefont
  {Bennenhei}, \citenamefont {Struve}, \citenamefont {Stephan}, \citenamefont
  {Kunte}, \citenamefont {Mitryakhin}, \citenamefont {Eilenberger},
  \citenamefont {Ohmer}, \citenamefont {Fischer}, \citenamefont {Silies},
  \citenamefont {Schneider},\ and\ \citenamefont
  {Esmann}}]{Bennenhei_OptMaterExpress2023}%
  \BibitemOpen
  \bibfield  {author} {\bibinfo {author} {\bibfnamefont {C.}~\bibnamefont
  {Bennenhei}}, \bibinfo {author} {\bibfnamefont {M.}~\bibnamefont {Struve}},
  \bibinfo {author} {\bibfnamefont {S.}~\bibnamefont {Stephan}}, \bibinfo
  {author} {\bibfnamefont {N.}~\bibnamefont {Kunte}}, \bibinfo {author}
  {\bibfnamefont {V.~N.}\ \bibnamefont {Mitryakhin}}, \bibinfo {author}
  {\bibfnamefont {F.}~\bibnamefont {Eilenberger}}, \bibinfo {author}
  {\bibfnamefont {J.}~\bibnamefont {Ohmer}}, \bibinfo {author} {\bibfnamefont
  {U.}~\bibnamefont {Fischer}}, \bibinfo {author} {\bibfnamefont
  {M.}~\bibnamefont {Silies}}, \bibinfo {author} {\bibfnamefont
  {C.}~\bibnamefont {Schneider}},\ and\ \bibinfo {author} {\bibfnamefont
  {M.}~\bibnamefont {Esmann}},\ }\bibfield  {title} {\bibinfo {title}
  {Polarized room-temperature polariton lasing in elliptical microcavities
  filled with fluorescent proteins},\ }\href
  {https://doi.org/10.1364/OME.496883} {\bibfield  {journal} {\bibinfo
  {journal} {Opt. Mater. Express}\ }\textbf {\bibinfo {volume} {13}},\ \bibinfo
  {pages} {2633} (\bibinfo {year} {2023})}\BibitemShut {NoStop}%
\bibitem [{\citenamefont {Nelsen}\ \emph {et~al.}(2013)\citenamefont {Nelsen},
  \citenamefont {Liu}, \citenamefont {Steger}, \citenamefont {Snoke},
  \citenamefont {Balili}, \citenamefont {West},\ and\ \citenamefont
  {Pfeiffer}}]{Nelsen_PRX2013}%
  \BibitemOpen
  \bibfield  {author} {\bibinfo {author} {\bibfnamefont {B.}~\bibnamefont
  {Nelsen}}, \bibinfo {author} {\bibfnamefont {G.}~\bibnamefont {Liu}},
  \bibinfo {author} {\bibfnamefont {M.}~\bibnamefont {Steger}}, \bibinfo
  {author} {\bibfnamefont {D.~W.}\ \bibnamefont {Snoke}}, \bibinfo {author}
  {\bibfnamefont {R.}~\bibnamefont {Balili}}, \bibinfo {author} {\bibfnamefont
  {K.}~\bibnamefont {West}},\ and\ \bibinfo {author} {\bibfnamefont
  {L.}~\bibnamefont {Pfeiffer}},\ }\bibfield  {title} {\bibinfo {title}
  {Dissipationless flow and sharp threshold of a polariton condensate with long
  lifetime},\ }\href {https://doi.org/10.1103/PhysRevX.3.041015} {\bibfield
  {journal} {\bibinfo  {journal} {Phys. Rev. X}\ }\textbf {\bibinfo {volume}
  {3}},\ \bibinfo {pages} {041015} (\bibinfo {year} {2013})}\BibitemShut
  {NoStop}%
\bibitem [{\citenamefont {Ardizzone}\ \emph {et~al.}(2022)\citenamefont
  {Ardizzone}, \citenamefont {Riminucci}, \citenamefont {Zanotti},
  \citenamefont {Gianfrate}, \citenamefont {Efthymiou-Tsironi}, \citenamefont
  {Su{\`a}rez-Forero}, \citenamefont {Todisco}, \citenamefont {De~Giorgi},
  \citenamefont {Trypogeorgos}, \citenamefont {Gigli}, \citenamefont {Baldwin},
  \citenamefont {Pfeiffer}, \citenamefont {Ballarini}, \citenamefont {Nguyen},
  \citenamefont {Gerace},\ and\ \citenamefont
  {Sanvitto}}]{Ardizzone_Nature2022}%
  \BibitemOpen
  \bibfield  {author} {\bibinfo {author} {\bibfnamefont {V.}~\bibnamefont
  {Ardizzone}}, \bibinfo {author} {\bibfnamefont {F.}~\bibnamefont
  {Riminucci}}, \bibinfo {author} {\bibfnamefont {S.}~\bibnamefont {Zanotti}},
  \bibinfo {author} {\bibfnamefont {A.}~\bibnamefont {Gianfrate}}, \bibinfo
  {author} {\bibfnamefont {M.}~\bibnamefont {Efthymiou-Tsironi}}, \bibinfo
  {author} {\bibfnamefont {D.~G.}\ \bibnamefont {Su{\`a}rez-Forero}}, \bibinfo
  {author} {\bibfnamefont {F.}~\bibnamefont {Todisco}}, \bibinfo {author}
  {\bibfnamefont {M.}~\bibnamefont {De~Giorgi}}, \bibinfo {author}
  {\bibfnamefont {D.}~\bibnamefont {Trypogeorgos}}, \bibinfo {author}
  {\bibfnamefont {G.}~\bibnamefont {Gigli}}, \bibinfo {author} {\bibfnamefont
  {K.}~\bibnamefont {Baldwin}}, \bibinfo {author} {\bibfnamefont
  {L.}~\bibnamefont {Pfeiffer}}, \bibinfo {author} {\bibfnamefont
  {D.}~\bibnamefont {Ballarini}}, \bibinfo {author} {\bibfnamefont {H.~S.}\
  \bibnamefont {Nguyen}}, \bibinfo {author} {\bibfnamefont {D.}~\bibnamefont
  {Gerace}},\ and\ \bibinfo {author} {\bibfnamefont {D.}~\bibnamefont
  {Sanvitto}},\ }\bibfield  {title} {\bibinfo {title} {Polariton
  {B}ose--{E}instein condensate from a bound state in the continuum},\ }\href
  {https://doi.org/10.1038/s41586-022-04583-7} {\bibfield  {journal} {\bibinfo
  {journal} {Nature}\ }\textbf {\bibinfo {volume} {605}},\ \bibinfo {pages}
  {447} (\bibinfo {year} {2022})}\BibitemShut {NoStop}%
\bibitem [{\citenamefont {Caputo}\ \emph {et~al.}(2018)\citenamefont {Caputo},
  \citenamefont {Ballarini}, \citenamefont {Dagvadorj}, \citenamefont
  {S{\'a}nchez Mu{\~{n}}oz}, \citenamefont {De Giorgi}, \citenamefont
  {Dominici}, \citenamefont {West}, \citenamefont {Pfeiffer}, \citenamefont
  {Gigli}, \citenamefont {Laussy}, \citenamefont {Szyma{\'{n}}ska},\ and\
  \citenamefont {Sanvitto}}]{Caputo_NatMat2018}%
  \BibitemOpen
  \bibfield  {author} {\bibinfo {author} {\bibfnamefont {D.}~\bibnamefont
  {Caputo}}, \bibinfo {author} {\bibfnamefont {D.}~\bibnamefont {Ballarini}},
  \bibinfo {author} {\bibfnamefont {G.}~\bibnamefont {Dagvadorj}}, \bibinfo
  {author} {\bibfnamefont {C.}~\bibnamefont {S{\'a}nchez Mu{\~{n}}oz}},
  \bibinfo {author} {\bibfnamefont {M.}~\bibnamefont {De Giorgi}}, \bibinfo
  {author} {\bibfnamefont {L.}~\bibnamefont {Dominici}}, \bibinfo {author}
  {\bibfnamefont {K.}~\bibnamefont {West}}, \bibinfo {author} {\bibfnamefont
  {L.~N.}\ \bibnamefont {Pfeiffer}}, \bibinfo {author} {\bibfnamefont
  {G.}~\bibnamefont {Gigli}}, \bibinfo {author} {\bibfnamefont {F.~P.}\
  \bibnamefont {Laussy}}, \bibinfo {author} {\bibfnamefont {M.}~\bibnamefont
  {Szyma{\'{n}}ska}},\ and\ \bibinfo {author} {\bibfnamefont {D.}~\bibnamefont
  {Sanvitto}},\ }\bibfield  {title} {\bibinfo {title} {Topological order and
  thermal equilibrium in polariton condensates},\ }\href
  {https://doi.org/10.1038/nmat5039} {\bibfield  {journal} {\bibinfo  {journal}
  {Nature Materials}\ }\textbf {\bibinfo {volume} {17}},\ \bibinfo {pages}
  {145} (\bibinfo {year} {2018})}\BibitemShut {NoStop}%
\bibitem [{\citenamefont {{Askitopoulos}}\ \emph {et~al.}(2019)\citenamefont
  {{Askitopoulos}}, \citenamefont {{Pickup}}, \citenamefont {{Alyatkin}},
  \citenamefont {{Zasedatelev}}, \citenamefont {{Lagoudakis}}, \citenamefont
  {{Langbein}},\ and\ \citenamefont {{Lagoudakis}}}]{Askitopoulos_Arxiv2019}%
  \BibitemOpen
  \bibfield  {author} {\bibinfo {author} {\bibfnamefont {A.}~\bibnamefont
  {{Askitopoulos}}}, \bibinfo {author} {\bibfnamefont {L.}~\bibnamefont
  {{Pickup}}}, \bibinfo {author} {\bibfnamefont {S.}~\bibnamefont
  {{Alyatkin}}}, \bibinfo {author} {\bibfnamefont {A.}~\bibnamefont
  {{Zasedatelev}}}, \bibinfo {author} {\bibfnamefont {K.~G.}\ \bibnamefont
  {{Lagoudakis}}}, \bibinfo {author} {\bibfnamefont {W.}~\bibnamefont
  {{Langbein}}},\ and\ \bibinfo {author} {\bibfnamefont {P.~G.}\ \bibnamefont
  {{Lagoudakis}}},\ }\bibfield  {title} {\bibinfo {title} {{Giant increase of
  temporal coherence in optically trapped polariton condensate}},\ }\href
  {https://doi.org/10.48550/arXiv.1911.08981} {\bibfield  {journal} {\bibinfo
  {journal} {arXiv e-prints}\ ,\ \bibinfo {eid} {arXiv:1911.08981}} (\bibinfo
  {year} {2019})},\ \Eprint {https://arxiv.org/abs/1911.08981}
  {arXiv:1911.08981 [cond-mat.quant-gas]} \BibitemShut {NoStop}%
\bibitem [{\citenamefont {Sigurdsson}\ \emph {et~al.}(2022)\citenamefont
  {Sigurdsson}, \citenamefont {Gnusov}, \citenamefont {Alyatkin}, \citenamefont
  {Pickup}, \citenamefont {Gippius}, \citenamefont {Lagoudakis},\ and\
  \citenamefont {Askitopoulos}}]{Sigurdsson_PRL2022}%
  \BibitemOpen
  \bibfield  {author} {\bibinfo {author} {\bibfnamefont {H.}~\bibnamefont
  {Sigurdsson}}, \bibinfo {author} {\bibfnamefont {I.}~\bibnamefont {Gnusov}},
  \bibinfo {author} {\bibfnamefont {S.}~\bibnamefont {Alyatkin}}, \bibinfo
  {author} {\bibfnamefont {L.}~\bibnamefont {Pickup}}, \bibinfo {author}
  {\bibfnamefont {N.~A.}\ \bibnamefont {Gippius}}, \bibinfo {author}
  {\bibfnamefont {P.~G.}\ \bibnamefont {Lagoudakis}},\ and\ \bibinfo {author}
  {\bibfnamefont {A.}~\bibnamefont {Askitopoulos}},\ }\bibfield  {title}
  {\bibinfo {title} {Persistent self-induced larmor precession evidenced
  through periodic revivals of coherence},\ }\href
  {https://doi.org/10.1103/PhysRevLett.129.155301} {\bibfield  {journal}
  {\bibinfo  {journal} {Phys. Rev. Lett.}\ }\textbf {\bibinfo {volume} {129}},\
  \bibinfo {pages} {155301} (\bibinfo {year} {2022})}\BibitemShut {NoStop}%
\bibitem [{\citenamefont {Gnusov}\ \emph {et~al.}(2021)\citenamefont {Gnusov},
  \citenamefont {Sigurdsson}, \citenamefont {T\"opfer}, \citenamefont
  {Baryshev}, \citenamefont {Alyatkin},\ and\ \citenamefont
  {Lagoudakis}}]{Gnusov_PRAppl2021}%
  \BibitemOpen
  \bibfield  {author} {\bibinfo {author} {\bibfnamefont {I.}~\bibnamefont
  {Gnusov}}, \bibinfo {author} {\bibfnamefont {H.}~\bibnamefont {Sigurdsson}},
  \bibinfo {author} {\bibfnamefont {J.}~\bibnamefont {T\"opfer}}, \bibinfo
  {author} {\bibfnamefont {S.}~\bibnamefont {Baryshev}}, \bibinfo {author}
  {\bibfnamefont {S.}~\bibnamefont {Alyatkin}},\ and\ \bibinfo {author}
  {\bibfnamefont {P.}~\bibnamefont {Lagoudakis}},\ }\bibfield  {title}
  {\bibinfo {title} {All-optical linear-polarization engineering in single and
  coupled exciton-polariton condensates},\ }\href
  {https://doi.org/10.1103/PhysRevApplied.16.034014} {\bibfield  {journal}
  {\bibinfo  {journal} {Phys. Rev. Appl.}\ }\textbf {\bibinfo {volume} {16}},\
  \bibinfo {pages} {034014} (\bibinfo {year} {2021})}\BibitemShut {NoStop}%
\bibitem [{\citenamefont {Alyatkin}\ \emph {et~al.}(2020)\citenamefont
  {Alyatkin}, \citenamefont {T\"opfer}, \citenamefont {Askitopoulos},
  \citenamefont {Sigurdsson},\ and\ \citenamefont
  {Lagoudakis}}]{Alyatkin_PhysRevLett124207402(2020)}%
  \BibitemOpen
  \bibfield  {author} {\bibinfo {author} {\bibfnamefont {S.}~\bibnamefont
  {Alyatkin}}, \bibinfo {author} {\bibfnamefont {J.~D.}\ \bibnamefont
  {T\"opfer}}, \bibinfo {author} {\bibfnamefont {A.}~\bibnamefont
  {Askitopoulos}}, \bibinfo {author} {\bibfnamefont {H.}~\bibnamefont
  {Sigurdsson}},\ and\ \bibinfo {author} {\bibfnamefont {P.~G.}\ \bibnamefont
  {Lagoudakis}},\ }\bibfield  {title} {\bibinfo {title} {Optical control of
  couplings in polariton condensate lattices},\ }\href
  {https://doi.org/10.1103/PhysRevLett.124.207402} {\bibfield  {journal}
  {\bibinfo  {journal} {Phys. Rev. Lett.}\ }\textbf {\bibinfo {volume} {124}},\
  \bibinfo {pages} {207402} (\bibinfo {year} {2020})}\BibitemShut {NoStop}%
\bibitem [{\citenamefont {Krantz}\ \emph {et~al.}(2019)\citenamefont {Krantz},
  \citenamefont {Kjaergaard}, \citenamefont {Yan}, \citenamefont {Orlando},
  \citenamefont {Gustavsson},\ and\ \citenamefont
  {Oliver}}]{KrantzetalQuantumEnginnersGuideAppPhysRev2019}%
  \BibitemOpen
  \bibfield  {author} {\bibinfo {author} {\bibfnamefont {P.}~\bibnamefont
  {Krantz}}, \bibinfo {author} {\bibfnamefont {M.}~\bibnamefont {Kjaergaard}},
  \bibinfo {author} {\bibfnamefont {F.}~\bibnamefont {Yan}}, \bibinfo {author}
  {\bibfnamefont {T.~P.}\ \bibnamefont {Orlando}}, \bibinfo {author}
  {\bibfnamefont {S.}~\bibnamefont {Gustavsson}},\ and\ \bibinfo {author}
  {\bibfnamefont {W.~D.}\ \bibnamefont {Oliver}},\ }\bibfield  {title}
  {\bibinfo {title} {{A quantum engineer's guide to superconducting qubits}},\
  }\href {https://doi.org/10.1063/1.5089550} {\bibfield  {journal} {\bibinfo
  {journal} {Applied Physics Reviews}\ }\textbf {\bibinfo {volume} {6}},\
  \bibinfo {pages} {021318} (\bibinfo {year} {2019})}\BibitemShut {NoStop}%
\bibitem [{\citenamefont {DiVincenzo}(2000)}]{DiVincenzo2000}%
  \BibitemOpen
  \bibfield  {author} {\bibinfo {author} {\bibfnamefont {D.~P.}\ \bibnamefont
  {DiVincenzo}},\ }\bibfield  {title} {\bibinfo {title} {The physical
  implementation of quantum computation},\ }\href@noop {} {\bibfield  {journal}
  {\bibinfo  {journal} {Fortschritte der Physik}\ }\textbf {\bibinfo {volume}
  {48}},\ \bibinfo {pages} {771} (\bibinfo {year} {2000})}\BibitemShut
  {NoStop}%
\bibitem [{\citenamefont {Nielsen}\ and\ \citenamefont
  {Chuang}(2010)}]{Book_nielsen_chuang_2010}%
  \BibitemOpen
  \bibfield  {author} {\bibinfo {author} {\bibfnamefont {M.~A.}\ \bibnamefont
  {Nielsen}}\ and\ \bibinfo {author} {\bibfnamefont {I.~L.}\ \bibnamefont
  {Chuang}},\ }\href {https://doi.org/10.1017/CBO9780511976667} {\emph
  {\bibinfo {title} {Quantum Computation and Quantum Information: 10th
  Anniversary Edition}}}\ (\bibinfo  {publisher} {Cambridge University Press},\
  \bibinfo {year} {2010})\BibitemShut {NoStop}%
\bibitem [{\citenamefont {Peng}\ \emph {et~al.}(2019)\citenamefont {Peng},
  \citenamefont {Hamerly}, \citenamefont {Soltani},\ and\ \citenamefont
  {Englund}}]{Peng_OptExpr2019}%
  \BibitemOpen
  \bibfield  {author} {\bibinfo {author} {\bibfnamefont {C.}~\bibnamefont
  {Peng}}, \bibinfo {author} {\bibfnamefont {R.}~\bibnamefont {Hamerly}},
  \bibinfo {author} {\bibfnamefont {M.}~\bibnamefont {Soltani}},\ and\ \bibinfo
  {author} {\bibfnamefont {D.~R.}\ \bibnamefont {Englund}},\ }\bibfield
  {title} {\bibinfo {title} {Design of high-speed phase-only spatial light
  modulators with two-dimensional tunable microcavity arrays},\ }\href
  {https://doi.org/10.1364/OE.27.030669} {\bibfield  {journal} {\bibinfo
  {journal} {Opt. Express}\ }\textbf {\bibinfo {volume} {27}},\ \bibinfo
  {pages} {30669} (\bibinfo {year} {2019})}\BibitemShut {NoStop}%
\bibitem [{\citenamefont {{Sedov}}\ \emph {et~al.}(2019)\citenamefont
  {{Sedov}}, \citenamefont {{Lukoshkin}}, \citenamefont {{Kalevich}},
  \citenamefont {{Hatzopoulos}}, \citenamefont {{Savvidis}},\ and\
  \citenamefont {{Kavokin}}}]{Sedov_arxiv2019}%
  \BibitemOpen
  \bibfield  {author} {\bibinfo {author} {\bibfnamefont {E.~S.}\ \bibnamefont
  {{Sedov}}}, \bibinfo {author} {\bibfnamefont {V.~A.}\ \bibnamefont
  {{Lukoshkin}}}, \bibinfo {author} {\bibfnamefont {V.~K.}\ \bibnamefont
  {{Kalevich}}}, \bibinfo {author} {\bibfnamefont {Z.}~\bibnamefont
  {{Hatzopoulos}}}, \bibinfo {author} {\bibfnamefont {P.~G.}\ \bibnamefont
  {{Savvidis}}},\ and\ \bibinfo {author} {\bibfnamefont {A.~V.}\ \bibnamefont
  {{Kavokin}}},\ }\bibfield  {title} {\bibinfo {title} {{Superfluid currents in
  half-moon polariton condensates}},\ }\href
  {https://doi.org/10.48550/arXiv.1910.00344} {\bibfield  {journal} {\bibinfo
  {journal} {arXiv e-prints}\ ,\ \bibinfo {eid} {arXiv:1910.00344}} (\bibinfo
  {year} {2019})},\ \Eprint {https://arxiv.org/abs/1910.00344}
  {arXiv:1910.00344 [cond-mat.mes-hall]} \BibitemShut {NoStop}%
\bibitem [{\citenamefont {Cherotchenko}\ \emph {et~al.}(2021)\citenamefont
  {Cherotchenko}, \citenamefont {Sigurdsson}, \citenamefont {Askitopoulos},\
  and\ \citenamefont {Nalitov}}]{Cherotchenko_PRB2021}%
  \BibitemOpen
  \bibfield  {author} {\bibinfo {author} {\bibfnamefont {E.~D.}\ \bibnamefont
  {Cherotchenko}}, \bibinfo {author} {\bibfnamefont {H.}~\bibnamefont
  {Sigurdsson}}, \bibinfo {author} {\bibfnamefont {A.}~\bibnamefont
  {Askitopoulos}},\ and\ \bibinfo {author} {\bibfnamefont {A.~V.}\ \bibnamefont
  {Nalitov}},\ }\bibfield  {title} {\bibinfo {title} {Optically controlled
  polariton condensate molecules},\ }\href
  {https://doi.org/10.1103/PhysRevB.103.115309} {\bibfield  {journal} {\bibinfo
   {journal} {Phys. Rev. B}\ }\textbf {\bibinfo {volume} {103}},\ \bibinfo
  {pages} {115309} (\bibinfo {year} {2021})}\BibitemShut {NoStop}%
\bibitem [{\citenamefont {Ma}\ \emph {et~al.}(2020{\natexlab{a}})\citenamefont
  {Ma}, \citenamefont {Berger}, \citenamefont {A{\ss}mann}, \citenamefont
  {Driben}, \citenamefont {Meier}, \citenamefont {Schneider}, \citenamefont
  {H{\"o}fling},\ and\ \citenamefont {Schumacher}}]{Ma_NatComm2020}%
  \BibitemOpen
  \bibfield  {author} {\bibinfo {author} {\bibfnamefont {X.}~\bibnamefont
  {Ma}}, \bibinfo {author} {\bibfnamefont {B.}~\bibnamefont {Berger}}, \bibinfo
  {author} {\bibfnamefont {M.}~\bibnamefont {A{\ss}mann}}, \bibinfo {author}
  {\bibfnamefont {R.}~\bibnamefont {Driben}}, \bibinfo {author} {\bibfnamefont
  {T.}~\bibnamefont {Meier}}, \bibinfo {author} {\bibfnamefont
  {C.}~\bibnamefont {Schneider}}, \bibinfo {author} {\bibfnamefont
  {S.}~\bibnamefont {H{\"o}fling}},\ and\ \bibinfo {author} {\bibfnamefont
  {S.}~\bibnamefont {Schumacher}},\ }\bibfield  {title} {\bibinfo {title}
  {Realization of all-optical vortex switching in exciton-polariton
  condensates},\ }\href {https://doi.org/10.1038/s41467-020-14702-5} {\bibfield
   {journal} {\bibinfo  {journal} {Nature Communications}\ }\textbf {\bibinfo
  {volume} {11}},\ \bibinfo {pages} {897} (\bibinfo {year}
  {2020}{\natexlab{a}})}\BibitemShut {NoStop}%
\bibitem [{\citenamefont {Cerna}\ \emph {et~al.}(2009)\citenamefont {Cerna},
  \citenamefont {Sarchi}, \citenamefont {Para\"{\i}so}, \citenamefont {Nardin},
  \citenamefont {L\'eger}, \citenamefont {Richard}, \citenamefont {Pietka},
  \citenamefont {El~Daif}, \citenamefont {Morier-Genoud}, \citenamefont
  {Savona}, \citenamefont {Portella-Oberli},\ and\ \citenamefont
  {Deveaud-Pl\'edran}}]{CernaPhysRevB80121309(2009)}%
  \BibitemOpen
  \bibfield  {author} {\bibinfo {author} {\bibfnamefont {R.}~\bibnamefont
  {Cerna}}, \bibinfo {author} {\bibfnamefont {D.}~\bibnamefont {Sarchi}},
  \bibinfo {author} {\bibfnamefont {T.~K.}\ \bibnamefont {Para\"{\i}so}},
  \bibinfo {author} {\bibfnamefont {G.}~\bibnamefont {Nardin}}, \bibinfo
  {author} {\bibfnamefont {Y.}~\bibnamefont {L\'eger}}, \bibinfo {author}
  {\bibfnamefont {M.}~\bibnamefont {Richard}}, \bibinfo {author} {\bibfnamefont
  {B.}~\bibnamefont {Pietka}}, \bibinfo {author} {\bibfnamefont
  {O.}~\bibnamefont {El~Daif}}, \bibinfo {author} {\bibfnamefont
  {F.}~\bibnamefont {Morier-Genoud}}, \bibinfo {author} {\bibfnamefont
  {V.}~\bibnamefont {Savona}}, \bibinfo {author} {\bibfnamefont {M.~T.}\
  \bibnamefont {Portella-Oberli}},\ and\ \bibinfo {author} {\bibfnamefont
  {B.}~\bibnamefont {Deveaud-Pl\'edran}},\ }\bibfield  {title} {\bibinfo
  {title} {Coherent optical control of the wave function of zero-dimensional
  exciton polaritons},\ }\href {https://doi.org/10.1103/PhysRevB.80.121309}
  {\bibfield  {journal} {\bibinfo  {journal} {Phys. Rev. B}\ }\textbf {\bibinfo
  {volume} {80}},\ \bibinfo {pages} {121309} (\bibinfo {year}
  {2009})}\BibitemShut {NoStop}%
\bibitem [{\citenamefont {Gnusov}\ \emph {et~al.}(2023)\citenamefont {Gnusov},
  \citenamefont {Harrison}, \citenamefont {Alyatkin}, \citenamefont {Sitnik},
  \citenamefont {Töpfer}, \citenamefont {Sigurdsson},\ and\ \citenamefont
  {Lagoudakis}}]{Gnusov_SciAdv2023}%
  \BibitemOpen
  \bibfield  {author} {\bibinfo {author} {\bibfnamefont {I.}~\bibnamefont
  {Gnusov}}, \bibinfo {author} {\bibfnamefont {S.}~\bibnamefont {Harrison}},
  \bibinfo {author} {\bibfnamefont {S.}~\bibnamefont {Alyatkin}}, \bibinfo
  {author} {\bibfnamefont {K.}~\bibnamefont {Sitnik}}, \bibinfo {author}
  {\bibfnamefont {J.}~\bibnamefont {Töpfer}}, \bibinfo {author} {\bibfnamefont
  {H.}~\bibnamefont {Sigurdsson}},\ and\ \bibinfo {author} {\bibfnamefont
  {P.}~\bibnamefont {Lagoudakis}},\ }\bibfield  {title} {\bibinfo {title}
  {Quantum vortex formation in the “rotating bucket” experiment with
  polariton condensates},\ }\href {https://doi.org/10.1126/sciadv.add1299}
  {\bibfield  {journal} {\bibinfo  {journal} {Science Advances}\ }\textbf
  {\bibinfo {volume} {9}},\ \bibinfo {pages} {eadd1299} (\bibinfo {year}
  {2023})}\BibitemShut {NoStop}%
\bibitem [{\citenamefont {del Valle-Inclan~Redondo}\ \emph
  {et~al.}(2023)\citenamefont {del Valle-Inclan~Redondo}, \citenamefont
  {Schneider}, \citenamefont {Klembt}, \citenamefont {Höfling}, \citenamefont
  {Tarucha},\ and\ \citenamefont {Fraser}}]{delValle_NanoLetters_2023}%
  \BibitemOpen
  \bibfield  {author} {\bibinfo {author} {\bibfnamefont {Y.}~\bibnamefont {del
  Valle-Inclan~Redondo}}, \bibinfo {author} {\bibfnamefont {C.}~\bibnamefont
  {Schneider}}, \bibinfo {author} {\bibfnamefont {S.}~\bibnamefont {Klembt}},
  \bibinfo {author} {\bibfnamefont {S.}~\bibnamefont {Höfling}}, \bibinfo
  {author} {\bibfnamefont {S.}~\bibnamefont {Tarucha}},\ and\ \bibinfo {author}
  {\bibfnamefont {M.~D.}\ \bibnamefont {Fraser}},\ }\bibfield  {title}
  {\bibinfo {title} {Optically driven rotation of exciton–polariton
  condensates},\ }\href {https://doi.org/10.1021/acs.nanolett.3c01021}
  {\bibfield  {journal} {\bibinfo  {journal} {Nano Letters}\ }\textbf {\bibinfo
  {volume} {23}},\ \bibinfo {pages} {4564} (\bibinfo {year}
  {2023})}\BibitemShut {NoStop}%
\bibitem [{\citenamefont {Sakurai}\ and\ \citenamefont
  {Napolitano}(2017)}]{Book_sakurai_napolitano_2017}%
  \BibitemOpen
  \bibfield  {author} {\bibinfo {author} {\bibfnamefont {J.~J.}\ \bibnamefont
  {Sakurai}}\ and\ \bibinfo {author} {\bibfnamefont {J.}~\bibnamefont
  {Napolitano}},\ }\href {https://doi.org/10.1017/9781108499996} {\emph
  {\bibinfo {title} {Modern Quantum Mechanics}}},\ \bibinfo {edition} {2nd}\
  ed.\ (\bibinfo  {publisher} {Cambridge University Press},\ \bibinfo {year}
  {2017})\BibitemShut {NoStop}%
\bibitem [{\citenamefont {Li}(2022)}]{QubitFidelity10.3389/fphy.2022.893507}%
  \BibitemOpen
  \bibfield  {author} {\bibinfo {author} {\bibfnamefont {K.}~\bibnamefont
  {Li}},\ }\bibfield  {title} {\bibinfo {title} {The qubit fidelity under
  different error mechanisms based on error correction threshold},\ }\bibfield
  {journal} {\bibinfo  {journal} {Frontiers in Physics}\ }\textbf {\bibinfo
  {volume} {10}},\ \href {https://doi.org/10.3389/fphy.2022.893507}
  {10.3389/fphy.2022.893507} (\bibinfo {year} {2022})\BibitemShut {NoStop}%
\bibitem [{\citenamefont {Horodecki}\ \emph {et~al.}(2009)\citenamefont
  {Horodecki}, \citenamefont {Horodecki}, \citenamefont {Horodecki},\ and\
  \citenamefont
  {Horodecki}}]{HorodeckiRevModPhys81865(2009)QuantumEntanglement}%
  \BibitemOpen
  \bibfield  {author} {\bibinfo {author} {\bibfnamefont {R.}~\bibnamefont
  {Horodecki}}, \bibinfo {author} {\bibfnamefont {P.}~\bibnamefont
  {Horodecki}}, \bibinfo {author} {\bibfnamefont {M.}~\bibnamefont
  {Horodecki}},\ and\ \bibinfo {author} {\bibfnamefont {K.}~\bibnamefont
  {Horodecki}},\ }\bibfield  {title} {\bibinfo {title} {Quantum entanglement},\
  }\href {https://doi.org/10.1103/RevModPhys.81.865} {\bibfield  {journal}
  {\bibinfo  {journal} {Rev. Mod. Phys.}\ }\textbf {\bibinfo {volume} {81}},\
  \bibinfo {pages} {865} (\bibinfo {year} {2009})}\BibitemShut {NoStop}%
\bibitem [{\citenamefont {Kavokin}\ \emph {et~al.}(2017)\citenamefont
  {Kavokin}, \citenamefont {Baumberg}, \citenamefont {Malpuech},\ and\
  \citenamefont {Laussy}}]{MicrocavitiesBook}%
  \BibitemOpen
  \bibfield  {author} {\bibinfo {author} {\bibfnamefont {A.~V.}\ \bibnamefont
  {Kavokin}}, \bibinfo {author} {\bibfnamefont {J.~J.}\ \bibnamefont
  {Baumberg}}, \bibinfo {author} {\bibfnamefont {G.}~\bibnamefont {Malpuech}},\
  and\ \bibinfo {author} {\bibfnamefont {F.~P.}\ \bibnamefont {Laussy}},\
  }\href {https://doi.org/10.1093/oso/9780198782995.001} {\emph {\bibinfo
  {title} {Microcavities}}},\ On Semiconductor Science and Technology\
  (\bibinfo  {publisher} {Oxford University Press},\ \bibinfo {year}
  {2017})\BibitemShut {NoStop}%
\bibitem [{\citenamefont {Clerk}\ \emph {et~al.}(2010)\citenamefont {Clerk},
  \citenamefont {Devoret}, \citenamefont {Girvin}, \citenamefont {Marquardt},\
  and\ \citenamefont {Schoelkopf}}]{ClerkQuantumNoiseRevModPhys821155(2010)}%
  \BibitemOpen
  \bibfield  {author} {\bibinfo {author} {\bibfnamefont {A.~A.}\ \bibnamefont
  {Clerk}}, \bibinfo {author} {\bibfnamefont {M.~H.}\ \bibnamefont {Devoret}},
  \bibinfo {author} {\bibfnamefont {S.~M.}\ \bibnamefont {Girvin}}, \bibinfo
  {author} {\bibfnamefont {F.}~\bibnamefont {Marquardt}},\ and\ \bibinfo
  {author} {\bibfnamefont {R.~J.}\ \bibnamefont {Schoelkopf}},\ }\bibfield
  {title} {\bibinfo {title} {Introduction to quantum noise, measurement, and
  amplification},\ }\href {https://doi.org/10.1103/RevModPhys.82.1155}
  {\bibfield  {journal} {\bibinfo  {journal} {Rev. Mod. Phys.}\ }\textbf
  {\bibinfo {volume} {82}},\ \bibinfo {pages} {1155} (\bibinfo {year}
  {2010})}\BibitemShut {NoStop}%
\bibitem [{\citenamefont {Burkard}\ \emph {et~al.}(2004)\citenamefont
  {Burkard}, \citenamefont {Koch},\ and\ \citenamefont
  {DiVincenzo}}]{BurkardMultiLevelDecoherencePhysRevB.69.064503(2004)}%
  \BibitemOpen
  \bibfield  {author} {\bibinfo {author} {\bibfnamefont {G.}~\bibnamefont
  {Burkard}}, \bibinfo {author} {\bibfnamefont {R.~H.}\ \bibnamefont {Koch}},\
  and\ \bibinfo {author} {\bibfnamefont {D.~P.}\ \bibnamefont {DiVincenzo}},\
  }\bibfield  {title} {\bibinfo {title} {Multilevel quantum description of
  decoherence in superconducting qubits},\ }\href
  {https://doi.org/10.1103/PhysRevB.69.064503} {\bibfield  {journal} {\bibinfo
  {journal} {Phys. Rev. B}\ }\textbf {\bibinfo {volume} {69}},\ \bibinfo
  {pages} {064503} (\bibinfo {year} {2004})}\BibitemShut {NoStop}%
\bibitem [{\citenamefont {Wootters}(1998)}]{WoottersPhysRevLett.80.2245(1998)}%
  \BibitemOpen
  \bibfield  {author} {\bibinfo {author} {\bibfnamefont {W.~K.}\ \bibnamefont
  {Wootters}},\ }\bibfield  {title} {\bibinfo {title} {Entanglement of
  formation of an arbitrary state of two qubits},\ }\href
  {https://doi.org/10.1103/PhysRevLett.80.2245} {\bibfield  {journal} {\bibinfo
   {journal} {Phys. Rev. Lett.}\ }\textbf {\bibinfo {volume} {80}},\ \bibinfo
  {pages} {2245} (\bibinfo {year} {1998})}\BibitemShut {NoStop}%
\bibitem [{\citenamefont {Johansson}\ \emph {et~al.}(2013)\citenamefont
  {Johansson}, \citenamefont {Nation},\ and\ \citenamefont
  {Nori}}]{QuTiP2JOHANSSON20131234}%
  \BibitemOpen
  \bibfield  {author} {\bibinfo {author} {\bibfnamefont {J.}~\bibnamefont
  {Johansson}}, \bibinfo {author} {\bibfnamefont {P.}~\bibnamefont {Nation}},\
  and\ \bibinfo {author} {\bibfnamefont {F.}~\bibnamefont {Nori}},\ }\bibfield
  {title} {\bibinfo {title} {Qutip 2: A python framework for the dynamics of
  open quantum systems},\ }\href
  {https://doi.org/https://doi.org/10.1016/j.cpc.2012.11.019} {\bibfield
  {journal} {\bibinfo  {journal} {Computer Physics Communications}\ }\textbf
  {\bibinfo {volume} {184}},\ \bibinfo {pages} {1234} (\bibinfo {year}
  {2013})}\BibitemShut {NoStop}%
\bibitem [{\citenamefont {Johansson}\ \emph {et~al.}(2012)\citenamefont
  {Johansson}, \citenamefont {Nation},\ and\ \citenamefont
  {Nori}}]{QuTip1JOHANSSON20121760}%
  \BibitemOpen
  \bibfield  {author} {\bibinfo {author} {\bibfnamefont {J.}~\bibnamefont
  {Johansson}}, \bibinfo {author} {\bibfnamefont {P.}~\bibnamefont {Nation}},\
  and\ \bibinfo {author} {\bibfnamefont {F.}~\bibnamefont {Nori}},\ }\bibfield
  {title} {\bibinfo {title} {Qutip: An open-source python framework for the
  dynamics of open quantum systems},\ }\href
  {https://doi.org/https://doi.org/10.1016/j.cpc.2012.02.021} {\bibfield
  {journal} {\bibinfo  {journal} {Computer Physics Communications}\ }\textbf
  {\bibinfo {volume} {183}},\ \bibinfo {pages} {1760} (\bibinfo {year}
  {2012})}\BibitemShut {NoStop}%
\bibitem [{\citenamefont {Ma}\ \emph {et~al.}(2020{\natexlab{b}})\citenamefont
  {Ma}, \citenamefont {Berger}, \citenamefont {A{\ss}mann}, \citenamefont
  {Driben}, \citenamefont {Meier}, \citenamefont {Schneider}, \citenamefont
  {H{\"o}fling},\ and\ \citenamefont {Schumacher}}]{Assmann1}%
  \BibitemOpen
  \bibfield  {author} {\bibinfo {author} {\bibfnamefont {X.}~\bibnamefont
  {Ma}}, \bibinfo {author} {\bibfnamefont {B.}~\bibnamefont {Berger}}, \bibinfo
  {author} {\bibfnamefont {M.}~\bibnamefont {A{\ss}mann}}, \bibinfo {author}
  {\bibfnamefont {R.}~\bibnamefont {Driben}}, \bibinfo {author} {\bibfnamefont
  {T.}~\bibnamefont {Meier}}, \bibinfo {author} {\bibfnamefont
  {C.}~\bibnamefont {Schneider}}, \bibinfo {author} {\bibfnamefont
  {S.}~\bibnamefont {H{\"o}fling}},\ and\ \bibinfo {author} {\bibfnamefont
  {S.}~\bibnamefont {Schumacher}},\ }\bibfield  {title} {\bibinfo {title}
  {Realization of all-optical vortex switching in exciton-polariton
  condensates},\ }\href {https://doi.org/10.1038/s41467-020-14702-5} {\bibfield
   {journal} {\bibinfo  {journal} {Nat. Commun.}\ }\textbf {\bibinfo {volume}
  {11}},\ \bibinfo {pages} {1} (\bibinfo {year}
  {2020}{\natexlab{b}})}\BibitemShut {NoStop}%
\bibitem [{\citenamefont {Yulin}\ \emph {et~al.}(2023)\citenamefont {Yulin},
  \citenamefont {Shelykh}, \citenamefont {Sedov},\ and\ \citenamefont
  {Kavokin}}]{yulin2023vorticity}%
  \BibitemOpen
  \bibfield  {author} {\bibinfo {author} {\bibfnamefont {A.~V.}\ \bibnamefont
  {Yulin}}, \bibinfo {author} {\bibfnamefont {I.~A.}\ \bibnamefont {Shelykh}},
  \bibinfo {author} {\bibfnamefont {E.~S.}\ \bibnamefont {Sedov}},\ and\
  \bibinfo {author} {\bibfnamefont {A.~V.}\ \bibnamefont {Kavokin}},\
  }\href@noop {} {\bibinfo {title} {Vorticity of polariton condensates in
  rotating traps}} (\bibinfo {year} {2023}),\ \Eprint
  {https://arxiv.org/abs/2306.17468} {arXiv:2306.17468 [physics.optics]}
  \BibitemShut {NoStop}%
\bibitem [{\citenamefont {Mair}\ \emph {et~al.}(2001)\citenamefont {Mair},
  \citenamefont {Vaziri}, \citenamefont {Weihs},\ and\ \citenamefont
  {Zeilinger}}]{ZeilingerNature412_313-316(2001)}%
  \BibitemOpen
  \bibfield  {author} {\bibinfo {author} {\bibfnamefont {A.}~\bibnamefont
  {Mair}}, \bibinfo {author} {\bibfnamefont {A.}~\bibnamefont {Vaziri}},
  \bibinfo {author} {\bibfnamefont {G.}~\bibnamefont {Weihs}},\ and\ \bibinfo
  {author} {\bibfnamefont {A.}~\bibnamefont {Zeilinger}},\ }\bibfield  {title}
  {\bibinfo {title} {{Entanglement of the orbital angular momentum states of
  photons}},\ }\href {https://doi.org/10.1038/35085529} {\bibfield  {journal}
  {\bibinfo  {journal} {Nature}\ }\textbf {\bibinfo {volume} {412}},\ \bibinfo
  {pages} {313–316} (\bibinfo {year} {2001})}\BibitemShut {NoStop}%
\bibitem [{\citenamefont {Plantenberg}\ and\ \citenamefont
  {Harmans}(2007)}]{PlantenbergNature2007}%
  \BibitemOpen
  \bibfield  {author} {\bibinfo {author} {\bibfnamefont {P.}~\bibnamefont
  {Plantenberg}, \bibfnamefont {J.and de~Groot}}\ and\ \bibinfo {author}
  {\bibfnamefont {C.~e.~a.}\ \bibnamefont {Harmans}},\ }\bibfield  {title}
  {\bibinfo {title} {{Demonstration of controlled-NOT quantum gates on a pair
  of superconducting quantum bits}},\ }\href
  {https://doi.org/10.1038/nature05896} {\bibfield  {journal} {\bibinfo
  {journal} {Nature}\ }\textbf {\bibinfo {volume} {447}},\ \bibinfo {pages}
  {836–839} (\bibinfo {year} {2007})}\BibitemShut {NoStop}%
\end{thebibliography}%

\end{document}